\documentclass{vldb}
\usepackage{graphicx}
\usepackage{multirow}
\usepackage{balance}  

\usepackage{times,amsmath,epsfig}
\usepackage{subcaption,url}
\usepackage{algorithm}
\usepackage{algpseudocode}

\begin{document}
	
\title{Rafiki: Machine Learning as an Analytics Service System  
}

\author{
	Wei Wang$^\dagger$, Sheng Wang$^\dagger$, Jinyang Gao$^\dagger$, Meihui Zhang$^\diamond$~\\ 
	Gang Chen$^\ddagger$, Teck Khim Ng$^\dagger$, Beng Chin Ooi$^\dagger$~\\
	\\
	 $^\dagger$National University of Singapore\\
	 $^\diamond$Beijing Institute of Technology \hspace{5mm} $^\ddagger$Zhejiang University \\
	 $^\dagger$\{wangwei, jinyang.gao, wangsh, ngtk, ooibc\}@comp.nus.edu.sg \\
	 $^\diamond$meihui\_zhang@bit.edu.cn, $^\ddagger$ cg@zju.edu.cn 
}

\maketitle

\begin{abstract}
Big data analytics is gaining massive momentum in the last few years.
Applying machine learning models to big data has become an implicit requirement or an expectation for most analysis tasks, especially on high-stakes applications.Typical applications include sentiment analysis against reviews for analyzing on-line products, image classification in food logging applications for monitoring user's daily intake and stock movement prediction. 
Extending traditional database systems to support the above analysis is intriguing but challenging. 
First, it is almost impossible to implement all machine learning models in the database engines. 
Second, expertise knowledge is required to optimize the training and inference procedures in terms of efficiency and effectiveness, which imposes heavy burden on the system users. 
In this paper, we develop and present a system, called Rafiki, to provide the training and inference service of machine learning models, and facilitate complex analytics on top of cloud platforms.
Rafiki provides distributed hyper-parameter tuning for the training service, and online ensemble modeling for the inference service which trades off between latency and accuracy. 
Experimental results confirm the efficiency, effectiveness, scalability and usability of Rafiki. 

\end{abstract}

\section{Introduction}\label{sec:intro}

\begin{figure}[htb]
	\centering
	\includegraphics[width=0.45\textwidth]{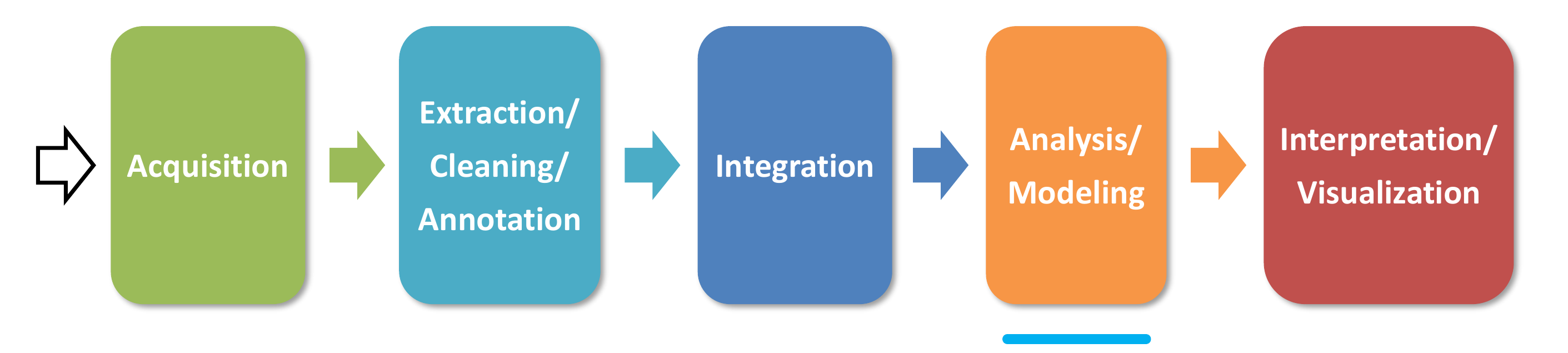}
	\caption{Data analytic pipeline.\label{fig:pipeline}}
\end{figure}

Data analysis plays an important role in extracting valuable insights from a huge amount of data. 
Database systems have been traditionally used for storing and analyzing structured data, spatial-temporal data, graph data, etc. 
Other data, such as multimedia data (e.g., images and free text), and domain specific data (e.g, medical data and sensor data), are being generated at fast speed and constitutes a significant portion of the Big Data~\cite{raey}. 
It is beneficial to analyze these data for database applications~\cite{wang2016database}. 
For instance, inferring the quality of a product from the review column in the sales database would help to explain the sales numbers; 
Analyzing food images from the food logging application can extract the food preference of people from different ages. 
However, the above analysis requires machine learning models, especially deep learning~\cite{lecun2015deep} models, for sentiment analysis~\cite{DBLP:journals/corr/abs-1103-0398} to classify the review as positive or negative, and image classification~\cite{DBLP:conf/nips/KrizhevskySH12} to recognize the food type from images. Figure~\ref{fig:pipeline} shows a pipeline of data analysis, where database systems have been widely used for the first 3 stages and machine learning is good at the 4th stage. 

One approach to integrating the machine learning techniques into database applications is to preprocess the data off-line and add the prediction results into a new column, e.g. a column for the food type. 
However, such off-line preprocessing has two-folds of restriction. First, it requires the developers to have expertise knowledge and experience of training machine learning models. Second, the queries cannot involve attributes of the object, e.g. the ingredients of food, if they are not predicted in the preprocessing step. 
In addition, it would waste a lot of time to do the prediction for all the rows if queries only read the food type column of few rows, e.g. due to a filtering on other columns.
Another approach is to carry out the prediction on-line as user-defined functions (UDFs) in the SQL query. 
However, it is challenging to implement all machine learning models in UDFs by database users~\cite{DBLP:conf/sigmod/ReABCJKR15}, for machine learning models vary a lot in terms of theory and implementation. 
It is also difficult to optimize the prediction accuracy in the database engine.

A better solution is to call the corresponding cloud machine learning service e.g. APIs, in the UDFs for each prediction (or analysis) task. 
Cloud service is economical, elastic and easy to use. 
With the resurgence of AI, cloud providers like Amazon (AWS), Google and Microsoft (Azure) have built machine learning services on their cloud platforms. 
There are two types of cloud services. The first one is to provide an API for each specific task, e.g. image classification and sentiment analysis. Such APIs are available on Amazon AWS\footnote{https://aws.amazon.com/machine-learning/} and Google cloud platform\footnote{https://cloud.google.com/products/machine-learning/}. 
The disadvantage is that the accuracy could be low since the models are trained by Amazon and Google with their data, which is likely to be different from the users' data. 
The second type of service overcomes this issue by providing the training service, where users can upload their own datasets to conduct training. 
This service is available on Amazon AWS and Microsoft Azure. However, only a limited number of machine learning models are supported~\cite{DBLP:journals/corr/ZhangZWZ17}. For example, only simple logistic regression or linear regression models\footnote{https://docs.aws.amazon.com/machine-learning/latest/dg/learning-algorithm.html} are supported by Amazon. Deep learning models such as convolutional neural networks (ConvNets) 
and recurrent neural networks (RNN) are not included. 

As a machine learning cloud service, it not only needs to cover a wide range of models, including deep learning models, but also provide an easy-to-use, efficient and effective service for users without much machine learning knowledge and experience, e.g. database users. Considering that different models may result in different performance in terms of efficiency (e.g. latency) and effectiveness (e.g. accuracy), the cloud service has to select the proper models for a given task. Moreover, most machine learning models and the training algorithms come with a set of hyper-parameters or knobs, e.g. number of layers and size of each layer in a neural network. Some hyper-parameters are vital for the convergence of the training algorithm and the final model performance, e.g. the learning rate of the stochastic gradient descent (SGD) algorithm. 
Manual hyper-parameter tuning requires rich experience and is tedious. 
Random search and Bayesian optimization are two popular automatic tuning approaches. 
However, both are costly as they have to train the model till convergence, which may take hundreds of hours. A distributed tuning platform is desirable. 
Besides training, inference\footnote{We use the term deployment, inference and serving interchangeably.} also matters as it directly affects the user experience. 
Machine learning products often ensemble multiple models and average the results to boost the prediction performance. However, ensemble modeling incurs a larger latency (i.e. response time) compared with using a single prediction model. Therefore, there is a trade-off between accuracy and latency.

There have been studies on these challenges for providing machine learning as a service. 
mlbench~\cite{DBLP:journals/corr/ZhangZWZ17} compares the cloud service of Amazon and Azure over a set of binary classification tasks. 
Ease.ml~\cite{mlease} builds a training platform with model selection aiming at optimizing the resource utilization. Google Vizier~\cite{46180} is a distributed hyper-parameter tuning platform that provides tuning service for other systems.  Clipper~\cite{201468} focuses on the inference by proposing a general framework and specific optimization for efficiency and accuracy. 

In this paper, we present a system, called Rafiki, to provide both the training and inference services for machine learning models. 
With Rafiki, (database) users are exempted from managing the hardware resource, constructing the (deep learning) models, tuning the hyper-parameters, optimizing the prediction accuracy and speed. 
Instead, they simply upload their datasets and configure the service to conduct training and then deploy the model for inference.
As a cloud service system~\cite{2002:PDS:876875.879015,relation-cloud}, Rafiki manages the hardware resources, failure recovery, etc. It comes with a set of built-in (deep learning) models for popular tasks such as image and text processing. In addition, we make the following contributions to make Rafiki easy-to-use, efficient and effective.

\begin{enumerate}
\item We propose a unified system architecture for both the training and the inference services. 
We observe that the two services share some common components such as a parameter server for model parameter storage, and distributed computing environment for distributed hyper-parameter tuning and parallel inference. By sharing the same underlying storage, communication protocols and computation resource, we implicitly avoid some technical debts~\cite{43146}. Moreover, by combining the two services together, Rafiki enables instant model deployment after training. 

\item For the training service, we first propose a general framework for distributed hyper-parameter tuning, which is extensible for popular hyper-parameter tuning algorithms including random search and Bayesian optimization. 
In addition, we propose a collaborative tuning scheme specifically for deep learning models, which uses the model parameters from the current top performing training trials to initialize new trials.

\item For the inference service, we propose a scheduling algorithm based on reinforcement learning to optimize the overall accuracy and reduce latency. Both algorithms are adaptive to the changes of the request arrival rate. 

\item We conduct micro-benchmark experiments to evaluate the performance of our proposed algorithms.

\end{enumerate}

In the reminder of this paper, we give the system architecture in Section~\ref{sec:sys}.  Section~\ref{sec:train} describes the training service and the distributed hyper-parameter tuning algorithm. The deployment service and optimization techniques are introduced in Section~\ref{sec:infer}. Section~\ref{sec:impl} describes the system implementation. We explain the experimental study in  Section~\ref{sec:exp} and then introduce related works in Section~\ref{sec:related}.

\section{Related Work}\label{sec:related}

Rafiki is a SaaS that provides analytics services based on machine learning. Optimization techniques are proposed for both the training (i.e. distributed hyper-parameter tuning) and the inference stage (i.e. accuracy and latency optimization). In this section, we review related work on SaaS, hyper-parameter tuning, inference optimization, and reinforcement learning which is adopted by Rafiki for inference optimization.

\subsection{Software as a Service}
Cloud computing has changed the way of IT operation in many companies by providing infrastructure as a service (IaaS), platform as a service (PaaS) and software as a service (SaaS). With IaaS, users and companies can use remote physical or virtual machines instead of establishing their own data center. Based on user requirements, IaaS can scale up the compute resources quickly. PaaS, e.g., Microsoft Azure, provides development toolkit and running environment, which can be adjusted automatically based on business demand. SaaS, including database as a service\cite{2002:PDS:876875.879015,relation-cloud}, installs software on cloud computing platforms and provides application services, which simplifies software maintenance. The `pay as you go' pricing model is convenient and economic for (small) companies and research labs.  

Recently, machine learning (especially deep learning) has gain a lot of interest due to its outstanding performance in analytic and predictive tasks. There are two primary steps to apply machine learning for an application, namely training and inference. Cloud providers, like Amazon AWS, Microsoft Azure and Google Cloud, have already included some services for the two steps. However, their training services have limited support for deep learning models, and their inference services cannot be customized for customers' data (see the explanation in Section~\ref{sec:intro}). There are also research papers towards efficient resource allocation for training~\cite{mlease} and efficient inference~\cite{201468} on the cloud. Rafiki differs from the existing cloud services on both the service types and the optimization techniques. Firstly, Rafiki allows users to train machine learning (including deep learning) models on their own data, and then deploy them for inference. Secondly, Rafiki has special optimization for the training and inference service, compared with the  existing approaches as explained in the following two subsections.

\subsection{Hyper-parameter Tuning}
To train a machine learning model for an application, we need to decide many hyper-parameters related to the model structure, optimization algorithm and data preprocessing operations. All hyper-parameter tuning algorithms work by empirical trials. Random search~\cite{Bergstra:2012:RSH:2188385.2188395} randomly selects the value of each hyper-parameter and then try it. It has shown to be more efficient than grid search that enumerates all possible hyper-parameter combinations. Bayesian optimization~\cite{2012arXiv1206.2944S} assumes the optimization function (hyper-parameters as the input and the inference performance as the output) follows Gaussian process. It samples the next point in the hyper-parameter space based on existing trials (according to an acquisition function). Reinforcement learning has recently been applied to tune the architecture related hyper-parameters~\cite{DBLP:journals/corr/ZophL16}. Google Vizier~\cite{46180} provides hyper-parameter tuning service on the cloud. Rafiki provides distributed hyper-parameter tuning service, which is compatible with all the above mentioned hyper-parameter tuning algorithms. In addition, it supports collaborative tuning that shares model parameters across trials. Our experiments confirm the effectiveness of the collaborative tuning scheme.

\subsection{Inference Optimization}
To improve the inference accuracy, ensemble modeling that trains and deploys multiple models is widely applied. For real-time inference, latency is another critical metric of the inference service. NoScope~\cite{Kang:2017:NON:3137628.3137664} proposes specific inference optimization for video querying. Clipper~\cite{201468} studies multiple optimization techniques to improve the throughput (via batch size) and reduce latency (via caching). It also proposes to do model selection for ensemble modeling using multi-armed bandit algorithms. Compared with Clipper, Rafiki provides both training and inference services, whereas Clipper focuses only on the inference service. Also, Clipper optimizes the throughput, latency and accuracy separately, whereas Rafiki models them together to find the optimal model selection and batch size. TensorFlow serving~\cite{199317} and TFX~\cite{46484} provide inference service for models trained using Tensorflow. Rafiki and Clipper are both implementation agnostic. They communicate with the training and inference programs in Docker containers via RPC.

\subsection{Reinforcement Learning}
In reinforcement learning (RL)\cite{Sutton:1998:IRL:551283}, each time an agent takes one action, it enters the next state. The environment returns a reward based on the action and the states. The learning objective is to maximize the aggregated rewards. RL algorithms decide the action to take based on the current state and experience gained from previous trials (exploration). Policy gradient based RL algorithms maximize the expected reward by taking actions following a policy function $\pi_\theta(a_t|s_t)$ over $n$ steps (Equation~\ref{eq:J}), where $\varsigma$ represents a trajectory of $n$  (action $a_t$, state $s_t$, reward $R_t$) tuples, and $\gamma$ is a decaying factor. The expectation is taken over all possible trajectories. $\pi_\theta$ is usually implemented using a multi-layer perceptron model ($\theta$ represents the parameters) that takes the state vector as input and generates the action (a scalar value for continuous action or a Softmax output for discrete actions). Equation~\ref{eq:J} is optimized using stochastic gradient ascent methods. Equation~\ref{eq:hatJ} is used as a `surrogate' function for the objective as its gradient is easy to compute via automatic differentiation.

\begin{eqnarray}
    J(\theta)&=&E_{\varsigma\sim\pi_\theta(\varsigma)}[\sum_{t=0}^n  \gamma^{t} R_{t} ],  \label{eq:J}\\
    \triangledown_\theta J(\theta)&=&E_{\varsigma\sim\pi_\theta(\varsigma)}[\sum_{t=0}^n \triangledown_\theta \log \pi_\theta(a_t|s_t) \sum_{t=0}^T  \gamma^{t} R_{t}] \label{eq:dJ}\\
    \hat{J}(\theta)&=&E_{\varsigma\sim\pi_\theta(\varsigma)}[\sum_{t=0}^T \log \pi_\theta(a_t|s_t) \sum_{t=0}^n  \gamma^{t} R_{t}] \label{eq:hatJ}
\end{eqnarray}

Many approaches have been proposed to improve the policy gradient by reducing the variance of the rewards, including actor-critic model~\cite{2016arXiv160201783M} which subtracts the reward $R_t$ by a baseline $V(s_t)$. $V(s_t)$ is an estimation of the reward, which is also approximated using a neural network like the policy function. Then $R_t$ in the above equations becomes $R_t - V(s_t)$. Yuxi~\cite{DBLP:journals/corr/Li17b} has done a survey of deep reinforcement learning, including the actor-critic model used in this paper.
\section{System Overview}\label{sec:sys}

\begin{figure*}[htb]
	\centering
	\includegraphics[width=0.98\textwidth]{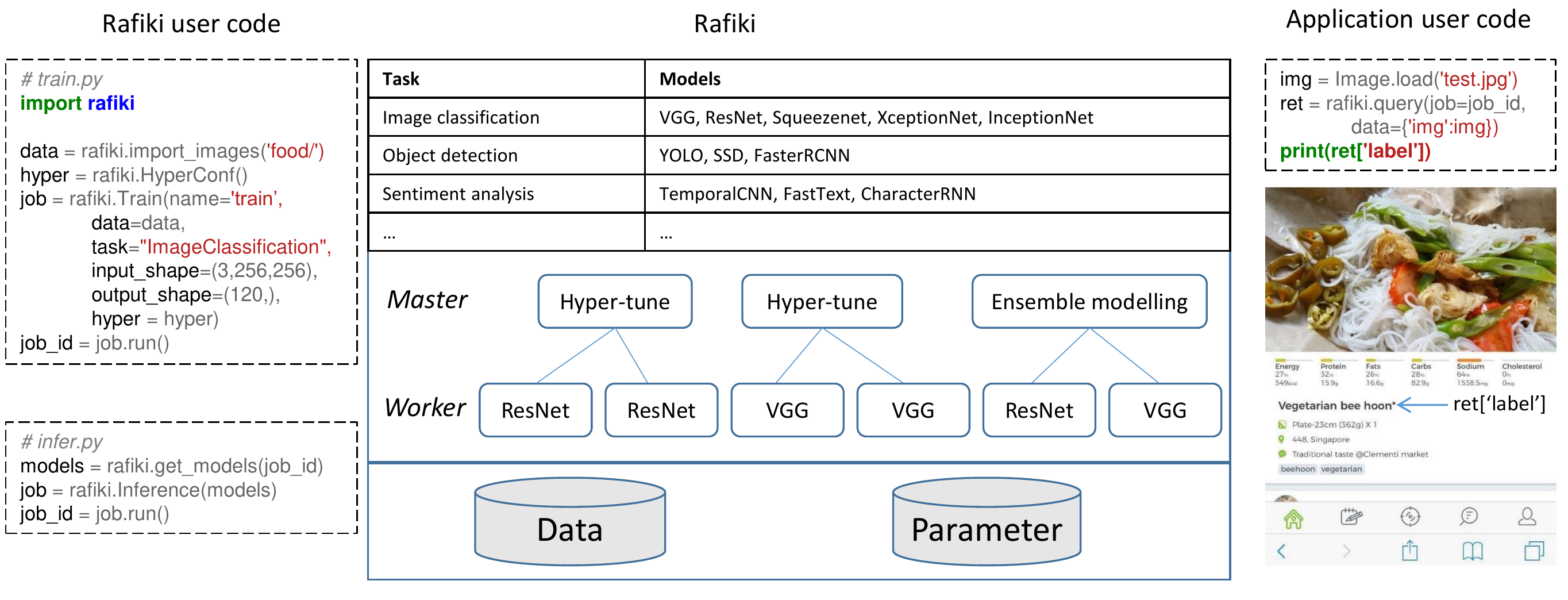}
	\caption{Overview of Rafiki.\label{fig:arch}}
\end{figure*}

In this section, we present the system architecture of Rafiki and illustrate its usability with some use cases. 

To use Rafiki, users simply configure the training or inference jobs through either RESTFul APIs or Python SDK. 
In Figure~\ref{fig:arch} (left column), we show one image classification example using the Python SDK. The training code consists of 4 lines. The first line loads the images from a folder into Rafiki's distributed data storage (HDFS), where all images from the same subfolder are labeled with the subfolder name, i.e. food name. 
The second line creates a configuration object representing hyper-parameter tuning options (see Section~\ref{sec:hyper} for details). The third line creates the training job by passing the data, the options and the input/output shapes (a tuple or a list of tuples). The input-output shapes are used for model customization. 
For example, ConvNets usually accept input images of a fixed shape, i.e. the input shape, and adapt the final output layer to generate the same number of outputs as the output shape, which could be the total number of classes or bounding-box shape. The last line submits the job through RESTFul APIs to Rafiki for execution. It returns a job ID as a handle for job monitoring. Once the training finishes, we can deploy the models instantly as shown in  \emph{infer.py}. It gets the model instances, each of which consists of the model name and the parameter names for retrieving the parameter values from Rafiki's parameter server. It then uses the second line to create the inference job with the given models. After the model is deployed by line 3, application users can submit their requests to this job for prediction as shown by the code in query.py. Applications like a mobile App can send RESTFul requests to do the query.

\begin{figure*}[htb]
\centering
\includegraphics[width=.98\textwidth]{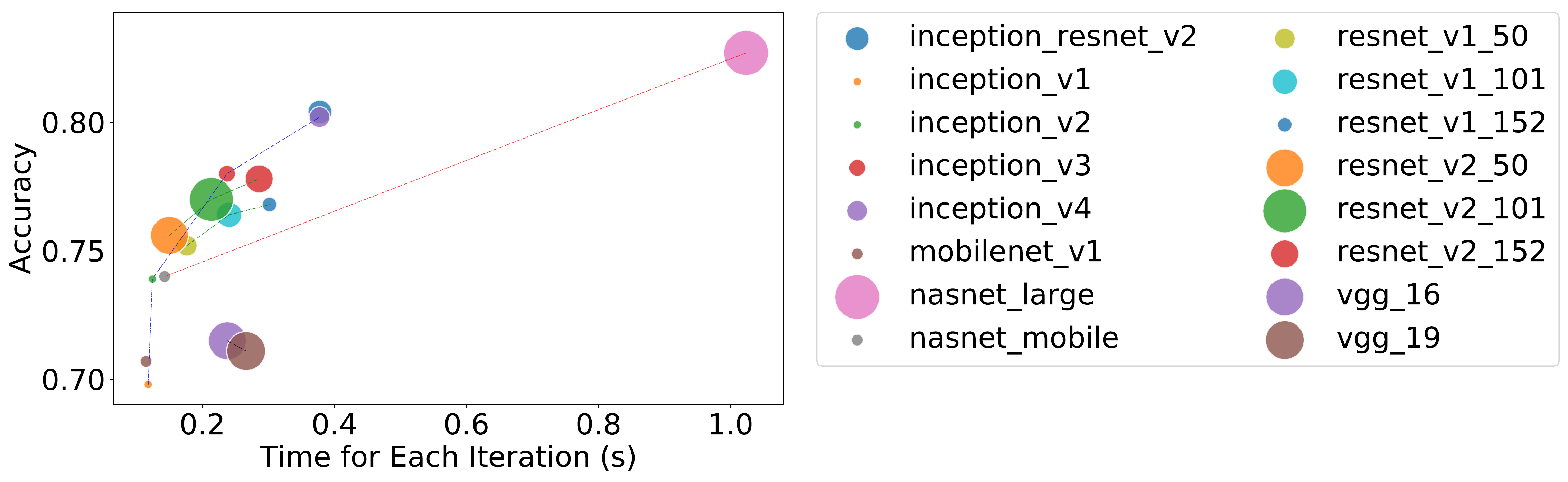}
\caption{Accuracy, inference time and memory footprint of popular ConvNets.\label{fig:convnets}}
\end{figure*}

For each training job, Rafiki selects the corresponding built-in models based on the task type. The table in Figure~\ref{fig:arch} lists some built-in tasks and the models. 
With the proliferation of machine learning, we are able to find open source implementations for almost every model. Figure~\ref{fig:convnets} compares the memory footprint, accuracy and inference speed of some open source ConvNets\footnote{https://github.com/tensorflow/models/tree/master/research/slim/}. 5 Groups of ConvNets are compared, including Inception ConvNets~\cite{DBLP:journals/corr/SzegedyLJSRAEVR14}, MobileNet~\cite{DBLP:journals/corr/HowardZCKWWAA17}, NASNets~\cite{DBLP:journals/corr/ZophL16}, ResNets~\cite{DBLP:journals/corr/HeZRS15} and VGGs~\cite{DBLP:journals/corr/SimonyanZ14a}.
The accuracy is measured based on the top-1 prediction of images from the validation dataset of ImageNet. 
The inference time and memory footprint is averaged over 50 iterations, each with 50 images (i.e. batch size=50).

 Hyper-parameter tuning (See Section~\ref{sec:hyper}) is conducted for every selected model. For example, two sets of hyper-parameters are being tested for ResNet~\cite{DBLP:journals/corr/HeZRS15} (resp. VGG~\cite{DBLP:journals/corr/SimonyanZ14a}) as shown in Figure~\ref{fig:arch}. 
The trained model parameters are stored in the global parameter server (PS), which is a persistent distributed in-memory storage. Once the training is done, the inference workers fetch the parameters from the PS efficiently. The inference job launches the models passed by user configuration. Ensemble modeling (See section~\ref{sec:infer}) is applied to optimize the inference accuracy of the requests from application (e.g. database) users.

Rafiki is compatible for any machine learning framework, including TensorFlow~\cite{199317}, Scikit-learn, Kears, etc., although we have built it on top of Apache Singa~\cite{mm15,singa,tomm} due to its efficiency and scalable architecture.
More details Rafiki implementation will be described in Section~\ref{sec:impl}.
\section{Training Service}\label{sec:train}

The training service consists of two steps, namely model selection and hyper-parameter tuning. 
We adopt a simple model selection strategy in Rafiki. For hyper-parameter tuning, we first propose a general programming model for distributed tuning, and then propose a collaborative tuning scheme.

\subsection{Model Selection}
Every built-in model in Rafiki is registered under a task, e.g. image classification or sentiment analysis. 
Each model is associated with some meta data including its training cost, e.g. the speed and memory consumption, and the performance on each dataset (e.g. accuracy or area-under-curve). 
Ease.ml~\cite{mlease} proposes to convert the model selection into a  multi-armed bandit problem, where every model (i.e. an arm) gets the chance of training. 
After many trials, the chance of under-performed models would be decreased.

We observe that the models for the same task perform consistently across datasets. For example, ResNet\cite{DBLP:journals/corr/HeZRS15} is better than AlexNet \cite{DBLP:conf/nips/KrizhevskySH12} and SqueezeNet~\cite{DBLP:journals/corr/IandolaMAHDK16} for a bunch of datasets including ImageNet~\cite{imagenet_cvpr09}, CIFAR-10\footnote{https://www.cs.toronto.edu/~kriz/cifar.html}, etc.. Therefore, we adopt a simple model selection strategy without any advanced analysis. 
We select the models with similar performance but with different architectures. This is to create a diverse model set~\cite{Kuncheva:2003:MDC:640211.640232} whose performance would be boosted when applying ensemble modeling.

\subsection{Distributed Hyper-parameter Tuning}\label{sec:hyper}

Hyper-parameter tuning usually has to run many different sets of hyper-parameters. Distributed tuning by running these instances in parallel is a natural solution for reducing the tuning time. There are three popular approaches for hyper-parameter tuning, namely grid search, random search~\cite{Bergstra:2012:RSH:2188385.2188395} and Bayesian optimization~\cite{7352306}. To support these algorithms, we need a general and extensible programming model.

\subsubsection{Programming Model}

\begin{table}
	\centering
	\caption{Hyper-parameter groups.\label{tb:hypers}}
	\begin{tabular}{|c|c|l|} \hline
		Group & Hyper-parameter&Example Domain\\ \hline
		\multirow{3}{*}{1. Data preprocessing} &  Image rotation & [0,30)\\ 
		& Image cropping & [0,32]\\
		& Whitening & \{PCA, ZCA\} \\\hline
		\multirow{3}{*}{2. Model architecture} & Number of layers  & $Z^+$\\
		& N\_cluster & $Z^+$\\
		& Kernel & \{Linear, RBF, Poly\}\\\hline 
		\multirow{3}{*}{3. Training algorithm} & Learning rate & $R^+$ \\
		&Weight decay & $R^+$\\
		&Momentum & $R^+$\\\hline 
		\end{tabular}
\end{table}

We cluster the hyper-parameters involved in training a machine learning model into three groups as shown in Table~\ref{tb:hypers}. Data preprocessing adopts different approaches to normalize the data, augment the dataset and extract features from the raw data, i.e. feature engineering. Most machine learning models have some tuning knobs which form the architecture's hyper-parameters, e.g. the number of trees in a random forest, number of layers of ConvNets  and the kernel function of SVM. The optimization algorithms, especially the gradient based optimization algorithms like SGD, have many hyper-parameters, including the initial learning rate, the decaying rate and decaying method (linear or exponential), etc. 

\begin{figure}[htb]
	\centering
	\includegraphics[width=0.49\textwidth]{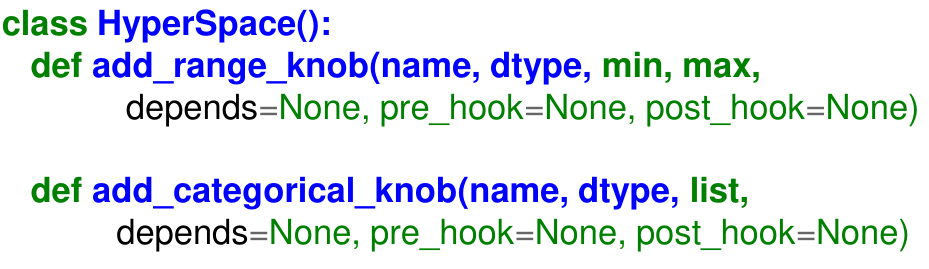}
	\caption{HyperSpace APIs.\label{fig:hyperspace}}
\end{figure}

From Table~\ref{tb:hypers} we can see that the  hyper-parameters could come from a range or a list of numbers or categories. All possible assignments of the hyper-parameters construct the \textbf{hyper-parameter space}, denoted as $\mathcal{H}$. Following the convention~\cite{46180}, we call one point in the space as a \textbf{trial}, denoted as $h$. Rafiki provides a \emph{HyperSpace} class with the functions shown in Figure~\ref{fig:hyperspace} for developers to specify the hyper-parameter space.

The first function defines the domain of a hyper-parameter as a range  $[min, max)$; \emph{dtype} represents the data type which could be float, integer or string. \emph{depends} is a list of other hyper-parameters whose values directly affect the generation of this hyper-parameter. For example, if the initial learning rate is very large, then we would prefer a larger decay rate to decease the learning rate quickly. Therefore, the decay rate has to be generated after generating the learning rate. To enforce such relation, we can add \emph{learning rate} into the \emph{depends} list and add a \emph{post\_hook} function to adjust the value of \emph{learning rate decay}.
The second function defines a categorical hyper-parameter, where \emph{list} represents the candidates. \emph{depends, pre\_hook, post\_hook} are analogous to those of the range knob.

The whole hyper-parameter tuning process for one model over a dataset is called a study. The performance of the trial $h$ is denoted as $p_h$. A larger $p_h$ indicates a better performance (e.g. accuracy).
For distributed hyper-parameter tuning, we have a master and multiple workers for each study. The master generates trials for workers, and the workers evaluate the trials. At one time, each worker trains the model with a given trial. 
The workflow of a study at the master side is explained in Algorithm~\ref{alg:master}. The master iterates over an event loop to collect the performance of each trial, i.e. $<h, p_h, t>$, and generates the next trial by \emph{TrialAdvisor} that implements the hyper-parameter search algorithm, e.g. random search or Bayesian optimization. It stops when there is no more trials to test or the user configured stop criteria is satisfied (e.g. total number of trials). Finally, the best parameters are put into the parameter server for the inference service.
The worker side keeps requesting trials from the master, conducting the training and reporting the results to the master.

\begin{algorithm}
	\caption{Study(HyperTune conf, TrialAdvisor adv)\label{alg:master}}
	\begin{algorithmic}[1]	
		\State num = 0	
		\While{conf.stop(num)}
			\State msg = ReceiveMsg()
			\If{msg.type == kRequest} 
				\State trial = adv.next(msg.worker)
				\If{trial is nil}		
					\State break
				\Else
					 \State send(msg.worker, trial)
				\EndIf
			\ElsIf{msg.type == kReport}
				\State adv.collect(msg.worker, msg.p, msg.trial)
			\ElsIf{msg.type == kFinish}
				\State num += 1
				\If{adv.is\_best(msg.worker)}
					\State send(msg.worker, kPut)
				\EndIf
			\EndIf		
		\EndWhile 
		\State return adv.best\_trial()
	\end{algorithmic}
\end{algorithm}

\subsubsection{Collaborative Tuning}
In the above section, we have introduced our distributed hyper-parameter tuning framework. Next, we extend the framework by proposing a collaborative tuning scheme.

SGD is widely used for training machine learning models. We often observe that the training loss stays in a plateau after a while, and then drops suddenly if we decrease the learning rate of SGD, e.g. from 0.1 to 0.01 and from 0.01 to 0.001\cite{DBLP:journals/corr/HeZRS15}.

Based on this observation, we can see that some hyper-parameters should be changed during training to get out of the plateau and derive better performance. Therefore, if we fix the model architecture and tune the hyper-parameters from group 1 and 3 in Table~\ref{tb:hypers}, then the model parameters trained using one hyper-parameter trial should be reused to initialize the same model with another trial. By selecting the parameters from the top performing trials to continue with the hyper-parameter tuning, the old trials are just like pre-training~\cite{Hinton:2006:FLA:1161603.1161605}. 
In fact, a good model initialization results in faster convergence and better performance\cite{Sutskever:2013:IIM:3042817.3043064}. Users typically fine-tune popular ConvNet architectures over their own datasets. Consequently, this collaborative tuning scheme is likely to converge to a good state. 

It is not straightforward to tune hyper-parameters related to the model architecture. This is because when the architecture changes, the parameters also change. For example, the parameters for a convolution layer with filter size 3x3 cannot be used to initialize the convolution layer of another ConvNet whose filter size is 5x5. However, during architecture tuning, there are many architectures available. It is likely that some architectures share the same configurations of one convolution layer. For instance, if ConvNet a's 3rd convolution layer and ConvNet b's 3rd layer have the same convolution setting,  then we can use the parameters $W$ from ConvNet a's 3rd layer to initialize ConvNet b's 3rd layer. We just store all $W$s in a parameter server and fetch the shape matched $W$ to initialize the layers in new trials (ConvNets). If the performance of the new trail is better than the older one, we overwrite the $W$ in the parameter server with the new values.

\begin{figure}[htb]
	\centering
	\includegraphics[width=0.47\textwidth]{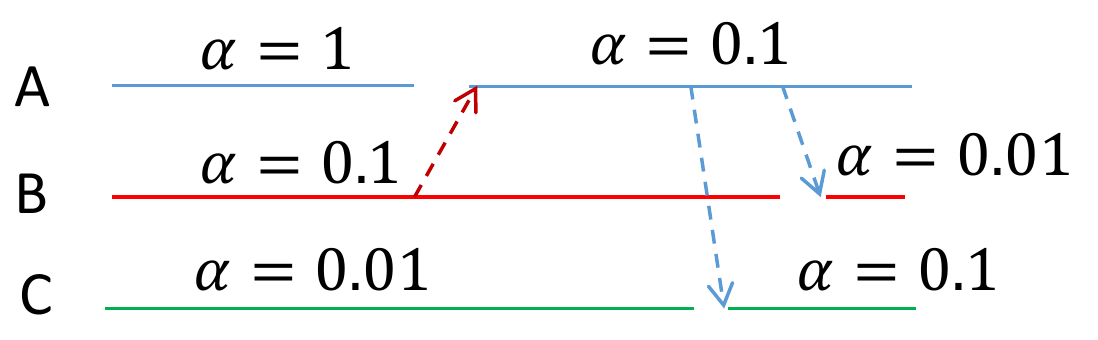}
	\caption{A collaborative tuning example.\label{fig:collaborative}}
\end{figure}

The whole process is illustrated in Figure~\ref{fig:collaborative}, where 3 workers are running trials to tune the hyper-parameters (e.g. the learning rate) of the same model. After a while, the performance of worker A stops increasing. The training stops automatically according to early stopping criteria, e.g the training loss is not decreasing for 5 consecutive epochs.  Early stopping is widely used for training machine learning models\footnote{\url{https://keras.io/callbacks/#earlystopping}}. The new trial on worker A uses the parameters from the current best worker, which is B. 
The work done by B serves as the pre-training for the new trial on A. Similarly, when C enters the plateau, the master instructs it to start a new trial on C using the parameters from A, whose model has the best performance at that point in time.

The control flow for this collaborative tuning scheme is described in Algorithm~\ref{alg:cmaster}. It is similar to Algorithm~\ref{alg:master} except that the master instructs the worker to save the model parameters into a global parameter server (Line 9) if the model's performance is significantly larger than the current best performance (Line 8). 
The performance difference, i.e. \emph{conf.delta} is a configuration parameter in \emph{HyperConf}, which is set according to the user's expectation about the performance of the model. 
For example, for MNIST image classification, an improvement of 0.1\% is very large as the current best performance is above 99\%. For CIFAR-10 image classification, we may set the delta to be 0.5\% as the best accuracy is about 97.4\%\cite{DBLP:journals/corr/abs-1708-04552}, which means that there is a bigger improvement space.

We notice that bad parameter initialization degrades the performance in our experiments. This is very serious for random search. Because the checkpoint from one trial with poor accuracy would affect the next trials as the model parameters are initialized into a poor state. To resolve this problem, a $\alpha$-greedy strategy is introduced in our implementation, which initializes the model parameters either by random initialization or from a pre-trained model checkpoint file. A threshold $\alpha$ represents the  probability of choosing random initialization and (1-$\alpha$) represents the probability of using pre-trained checkpoint files. $\alpha$ decreases gradually to decrease the chance of random initialization, i.e. increasing the chance of CoStudy. This $\alpha$-greedy strategy is widely used in reinforcement learning to balance the exploration and exploitation. 


\begin{algorithm}
	\caption{CoStudy(HyperTune conf, TrialAdvisor adv)\label{alg:cmaster}}
	\begin{algorithmic}[1]	
		\State num = 0, best\_p = 0
		\While{conf.stop(num)}
			\State msg = ReceiveMsg()
			\If{msg.type == kRequest} 
				\State ... \Comment{// same as Algorithm~\ref{alg:master}}
			\ElsIf{msg.type == kReport}
				\State adv.collect(msg.worker, msg.p, msg.trial)
				\If{msg.p - best\_p $>$ conf.delta }
					\State send(msg.worker, kPut)
					\State best\_p = msg.p
				\ElsIf{adv.early\_stopping(msg.worker, conf)}
					\State send(msg.worker, kStop)
				\EndIf				
			\ElsIf{msg.type == kFinish}
					\State num += 1
			\EndIf		
		\EndWhile 
		\State return adv.best\_trial()
	\end{algorithmic}
\end{algorithm}

\section{Inference Service}\label{sec:infer}

Inference service provides real-time request serving by deploying the trained model.
Other services, like database services, simply optimize the throughput with the constraint on latency, which is set manually as a service level objective (SLO), denoted as $\tau$, e.g. $\tau=0.1$ seconds. 
For machine learning services, accuracy becomes an important optimization objective. The accuracy refers to a wide range of performance measurements, e.g. negative error, precision, recall, F1, area under curve, etc. A larger value (accuracy) indicates better performance. If we set latency as a hard constraint as shown in Equation~\ref{eq:opt}, overdue requests would get `time out' responses. 
Typically, a delayed response is better than an error of `time out' for the end-users. Hence, we process the requests in the queue sequentially following FIFO (first-in-first-out). 
The objective is to maximize the accuracy and minimize the exceeding time according to $\tau$. 

However, typically, there is a trade-off between accuracy and latency. 
For example, ensemble modeling with more models increases both the accuracy and the latency. 
We shall optimize the model selection for ensemble modeling in Section~\ref{sec:multiple}.
Before that, we discuss a simpler case with a single inference model. Table~\ref{tb:notation} summarizes the notations used in this section.

\begin{eqnarray}
&\max Accuracy(S) \label{eq:opt}\\
&\text{subject to} \ \forall s\in S, l(s) < \tau \nonumber 
\end{eqnarray}

\begin{table}
	\centering
	\caption{Notations.\label{tb:notation}}
	\begin{tabular}{|l|l|}
		\hline 
		Name & Definition\\ \hline
		$S$ & request list  \\ \hline 
		$M$ & model list  \\ \hline 
		$\tau$ & latency requirement \\ \hline 
		$b \in B$ & one batch size from a candidate list \\ \hline 
	    $q_k$ & the $k-$th oldest requests in the queue \\ \hline 
	    $q_{:k}$ & is the oldest $k$ requests \\ \hline 
	    $q_{k:}$ & is the latest $|Q|-k$ requests \\ \hline 
	    $c(b)$ & inference time for batch size b \\ \hline
	    $c(m, b)$ & inference time for model $m$ and batch size $b$ \\ \hline 
	    $w(s)$ & waiting time for a request s in the queue \\ \hline 
	    $l(s)$ & latency (waiting + inference time) of a request \\ \hline 
	    $\beta$ & balancing factor between accuracy and latency \\ \hline 
	    $\mathbf{v}$ & binary vector for model selection \\ \hline 
	    $R()$ & reward function over a set of requests \\ \hline 
	\end{tabular}
\end{table}

\subsection{Single Inference Model}\label{sec:single}
When there is only one single model deployed for an application, the accuracy of the inference service is fixed from the system's perspective. Therefore, the optimization objective is reduced to minimizing the exceeding time, which is formalized in Equation~\ref{eq:single}.

\begin{eqnarray}\label{eq:single}
\min \frac{\sum_{s\in S} max(0, l(s)-\tau) }{|S|}
\end{eqnarray}

The latency $l(s)$ of a request includes the waiting time in the queue $w(s)$, and the inference time which depends on the model complexity, hardware efficiency (i.e. FLOPS) and the batch size. The batch size decides the number of requests to be processed together. 
Modern processing units, like CPU and GPU, exploit data parallelism techniques (e.g. SIMD) to improve the throughput and reduce the computation cost. Hence, a large batch size is necessary to saturate the parallelism capacity.
Once the model is deployed on a cloud platform, the model complexity and hardware efficiency are fixed. Therefore, Rafiki tunes the batch size to optimize the latency. 

To construct a large batch, e.g. with $b$ requests, we have to delay the processing until all $b$ requests arrive, which may incur a large latency for the old requests if the request arrival rate is low. The optimal batch size is thus influenced by SLO $\tau$, the queue status (e.g. the waiting time), and the request arrival rate which varies along time. Since the inference time of two similar batch sizes varies little, e.g. b=8 and b=9, a candidate batch size list should include values that have significant difference with each other w.r.t
the inference time, e.g. $B=\{16, 32, 48, 64,...\}$. 
The largest batch size is determined by the system (GPU) memory.
$c(b)$, the latency of processing a batch of b requests $b\in B$, is determined by the hardware resource (e.g. GPU memory), and the model's complexity. Figure~\ref{fig:convnets} shows the inference time, memory footprint and accuracy of popular ConvNets trained on ImageNet.

Algorithm~\ref{alg:greedy} shows a greedy solution for this problem. 
It always applies the largest batch size possible. If the queue length (i.e. number of requests in the queue) is larger than the largest batch size $b = \max(B)$, then the oldest $b$ requests are processed in one batch. Otherwise, it waits until the oldest request ($q_0$) is about to overdue as checked by Line 8. $b$ is the largest 
batch size in $B$ that is smaller or equal to the queue length (Line 8). $\delta$ is a back-off constant, which is equivalent to reducing the batch size in Additive-Increase-Multiplicative-Decrease scheme (AIMD)\cite{201468}, e.g. $\delta=0.1\tau$.

\begin{algorithm}
	\caption{Inference(Queue q, Model m)\label{alg:greedy}}
	\begin{algorithmic}[1]	
		\While{True}
		    \State $b = \max B$
			\If{$len(q)>=b$}
			    \State m.infer($q_{0:b}$)
			    \State deque($q_{0:b}$)
			\Else
                \State $b = \max \{b\in B, b<=len(q)\}$			    
			    \If{$c(b) + w(q_0) + \delta >= \tau$} 
			        \State m.infer($q_{0:b}$)
			        \State deque($q_{0:b}$)
			    \EndIf
			\EndIf    
		\EndWhile 
	\end{algorithmic}
\end{algorithm}

\subsection{Multiple Inference Models} \label{sec:multiple}

\begin{figure}[htb]
\centering
\includegraphics[width=.48\textwidth]{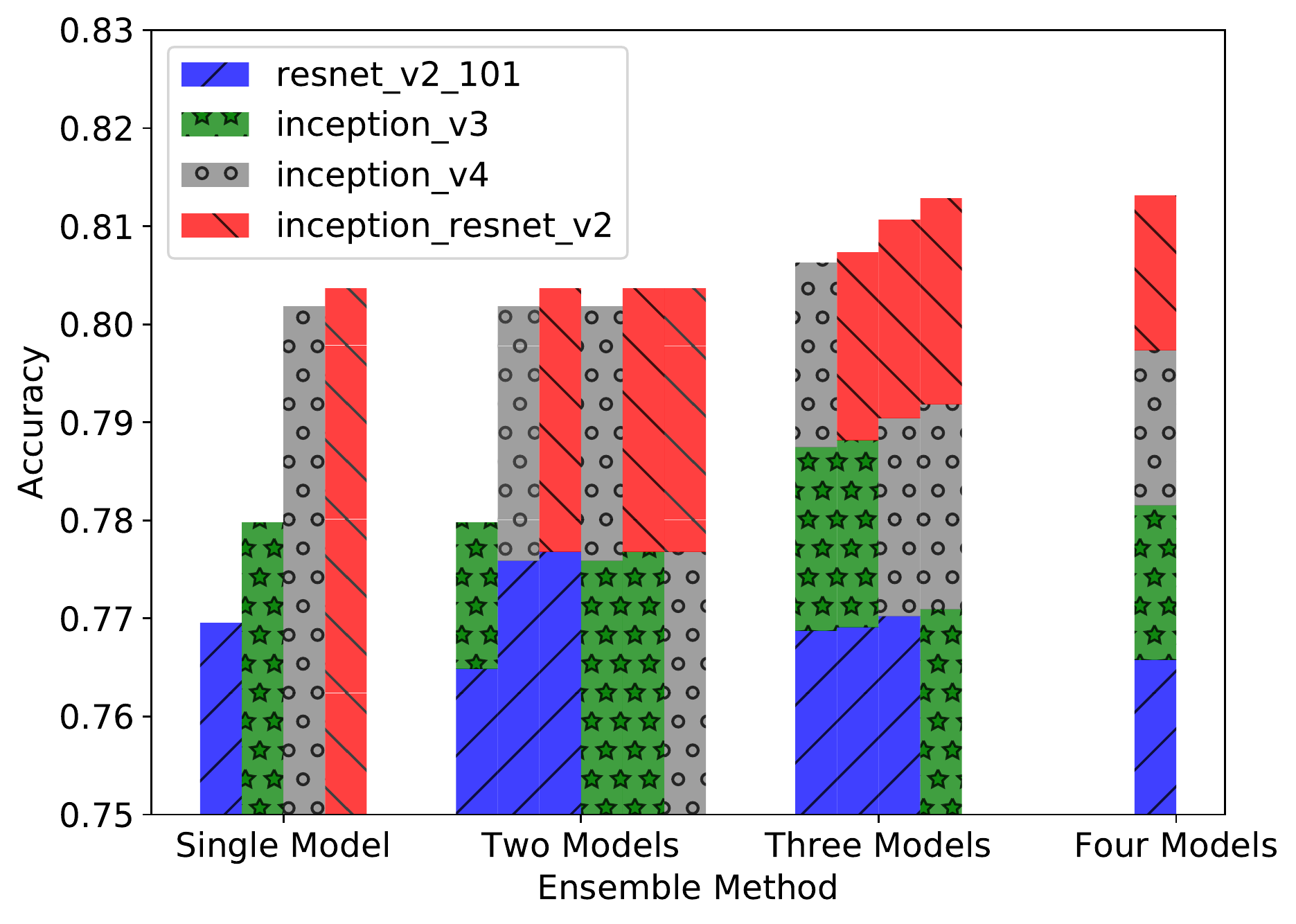}
\caption{Accuracy of ensemble modeling with different models.\label{fig:ensemble}}
\end{figure}

Ensemble modeling is an effective and popular approach to improve the inference accuracy. To give an example, we compare the performance of different ensemble of 4 ConvNets as shown in Figure~\ref{fig:ensemble}.  Majority voting is applied to aggregate the predictions. The accuracy is evaluated over the validation dataset of ImageNet. Generally with more models, the accuracy is better. The exception is that the ensemble of resnet\_v2\_101 and inception\_v3, which is not as good as the single best model, i.e. inception\_resnet\_v2. In fact, the prediction of the ensemble modeling is the same as inception\_v3 because when there is a tie, the prediction from the model with the best accuracy is selected as the final prediction.

Parallel ensemble modeling by running one model per node (or GPU) is a straight-forward way to scale the system and improve the throughput. However, the latency could be high due to stragglers. For example, as shown in Figure~\ref{fig:convnets}, the node running \emph{nasnet\_large} would be very slow although its accuracy is high. In addition, ensemble modeling is also costly in terms of throughput when compared with serving different requests on different nodes (i.e. no ensemble). We can see that there is a trade-off between latency (throughput) and accuracy, which is controlled by the model selection. If the requests arrive slowly, we simply run all models for each batch to get the best accuracy. However, when the request arrival rate is high, like in Section~\ref{sec:single}, we have to select different models for different requests to increase the throughput and reduce the latency. In addition, the model selection for the current batch also affects the next batch. For example, if we use all models for a batch, the next batch has to wait until at least one model finishes. 

To solve Equation~\ref{eq:opt}, we have to balance the accuracy and latency to get a single optimization objective. 
We move the latency term into the objective as shown in Equation~\ref{eq:multi-reward}. It maximizes a reward function $R$ related to the prediction accuracy and penalizes overdue requests. $\beta$ balances the accuracy and the latency in the objective. If the ground truth of each request is not available for evaluating the accuracy, which is the normal case, we have to find a surrogate accuracy.

\begin{eqnarray}
\max R(S) - \beta R(\{s\in S, l(s) > \tau\})  \label{eq:multi-reward}
\end{eqnarray}

Like the analysis for single inference model, we need to consider the batch size selection as well. It is difficult to design an optimal policy for this complex decision making problem, which decides both the model selection and batch size selection. In this paper, we propose to optimize Equation~\ref{eq:multi-reward} using reinforcement learning (RL). RL optimizes an objective over a long term by trying different actions and entering the corresponding states to collect rewards. By setting the reward as Equation~\ref{eq:multi-reward} and defining the actions to be model selection and batch selection, we can apply RL for our optimization problem.

RL has three core concepts, namely, the action, reward and state. Once these three concepts are defined, we can apply existing RL algorithms to do optimization. We define the three concepts w.r.t our optimization problem as follows.
First, the state space consists of : a) the queue status represented by the waiting time of each request in the queue. The waiting time of all requests form a feature vector. To generate a fixed length feature vector, we pad with 0 for the shorter queues and truncate the longer queues. 
b) the model status represented by a vector including the inference time for different models with different batch sizes, i.e. $c(m, b), m\in M, b \in B$, and the left time to finish the existing requests dispatched to it.  The two feature vectors are concatenated into a state feature vector, which is the input to the RL model for generating the action.
Second, the \emph{action} decides the batch size $b$ and model selection represented by a binary vector $\mathbf{v}$ of length $|M|$ (1 for selected; 0 for unselected). The action space size is thus $(2^{|M|}-1)*|B|$. We exclude the case where $\mathbf{v}=\mathbf{0}$, i.e. none of the models are selected. Third, following Equation~\ref{eq:multi-reward}, the \emph{reward} for one batch of requests without ground truth labels is defined in Equation~\ref{eq:multi_acc_reward},  $a(M[\mathbf{v}])$ is the accuracy of the selected models (ensemble modeling). In this paper, we use the accuracy evaluated on a validation dataset as the surrogate accuracy. In the experiment, we use the ImageNet's validation dataset to evaluate the accuracy of different ensemble combinations for image classification. The results are shown in Figure~\ref{fig:ensemble}.
The reward shown in Equation~\ref{eq:multi_acc_reward} considers the accuracy, the latency (indirectly represented by the number of overdue requests) and the number of requests. 
\begin{eqnarray}\label{eq:multi_acc_reward}
a(M[\mathbf{v}]) * (b - \beta |\{s\in \text{batch}| l(s)>\tau\}|)
\end{eqnarray}

With the state, action and reward well defined, we apply the actor-critic algorithm~\cite{DBLP:journals/corr/SchulmanWDRK17} to optimize the overall reward by learning a good policy for selecting the models and batch size.

\section{System Implementation}\label{sec:impl}

In this section, we introduce the implementation details of Rafiki, including the cluster management, data and parameter storage, and failure recovery. 

\subsection{Cluster Management}

\begin{figure}[htb]
	\centering
	\includegraphics[width=0.4\textwidth]{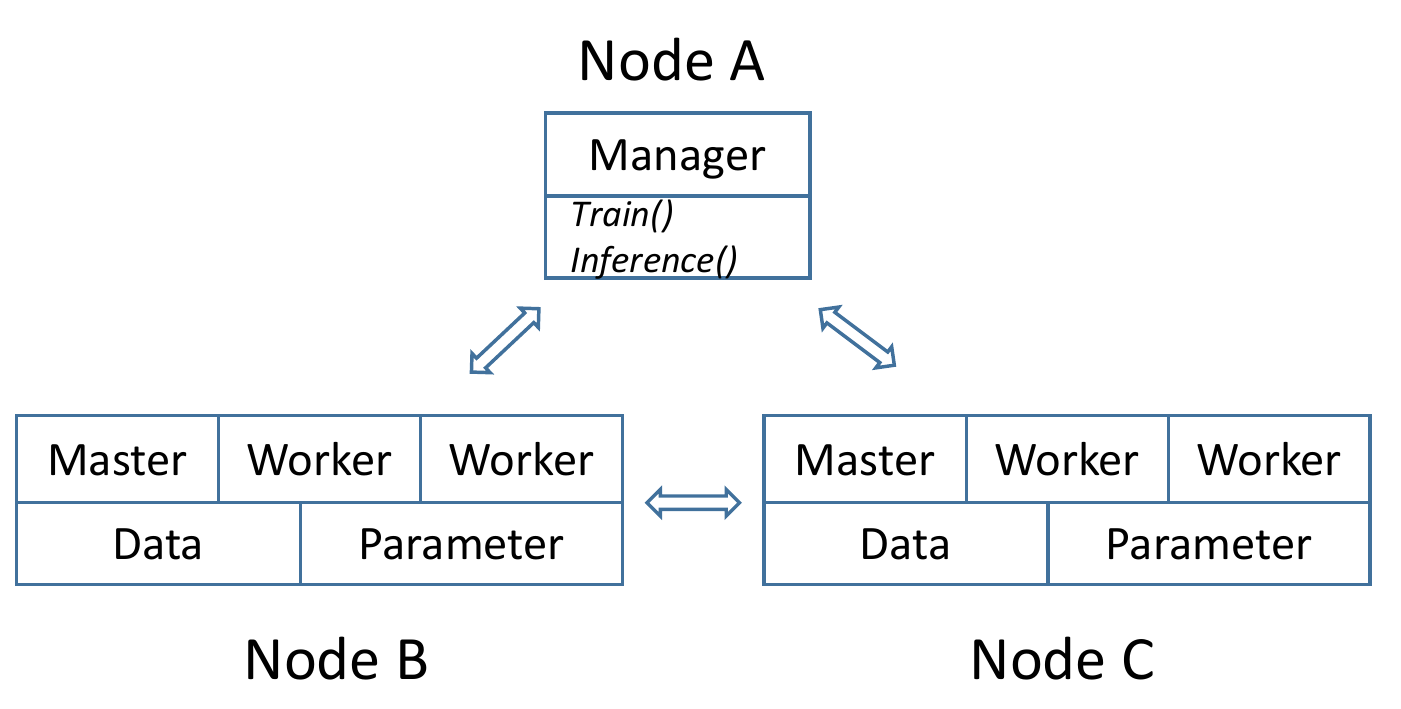}
	\caption{Rafiki cluster topology.\label{fig:cluster}}
\end{figure}

Rafiki uses Kubernetes to manage the Docker containers, which run the masters, workers, data servers and parameter servers as shown in Figure~\ref{fig:cluster}. 
Docker container is widely used for application deployment, which bundles the application code (e.g. training and inference code) and libraries (e.g. Keras) to avoid tedious environment setup. New models, hyper-parameter tuning algorithms, and ensemble modeling approaches are deployed as new Docker containers.
On each physical node, there could be multiple masters and workers for different jobs, including both training and inference jobs. These jobs are started by the Rafiki manager, which accepts job submissions from users (See Figure~\ref{fig:arch}). 
Once a job is launched, users  communicate with its master directly to get the training progress or submit query requests. 
Rafiki prefers to locate the master and workers for the same job in the same physical node to avoid network communication overhead.

\subsection{Data and Parameter Storage}

Deep learning models are typically trained over large datasets that would consume a lot of space if stored in CPU memory. 
Therefore,  
Rafiki uses HDFS for data storage, where the `data nodes'  are also docker containers. 
Users upload their datasets into the HDFS via Rafiki utility functions, e.g. $rafiki.import\_images$ (see Figure~\ref{fig:arch}). The training dataset is downloaded to a local directory before training, via $rafiki.download()$ by passing the dataset name.

For model parameters, Rafiki has its own distributed parameter server. The parameter server is designed with special optimization for hyper-parameter training and inference. In particular, the hyper-parameters will be cached in memory if they are accessed frequently, e.g. when Rafiki is doing hyper-parameter training. Otherwise, they are stored in HDFS. 
The parameters trained for the same model but different datasets can be shared as long as the privacy setting is public. It has been shown that training warm-up by using the parameters pre-trained on other datasets speeds up the training~\cite{46484}.

\begin{figure*}[h!]
	\centering
	\begin{subfigure}{0.43\textwidth}
		\centering
		\includegraphics[width=0.9\textwidth]{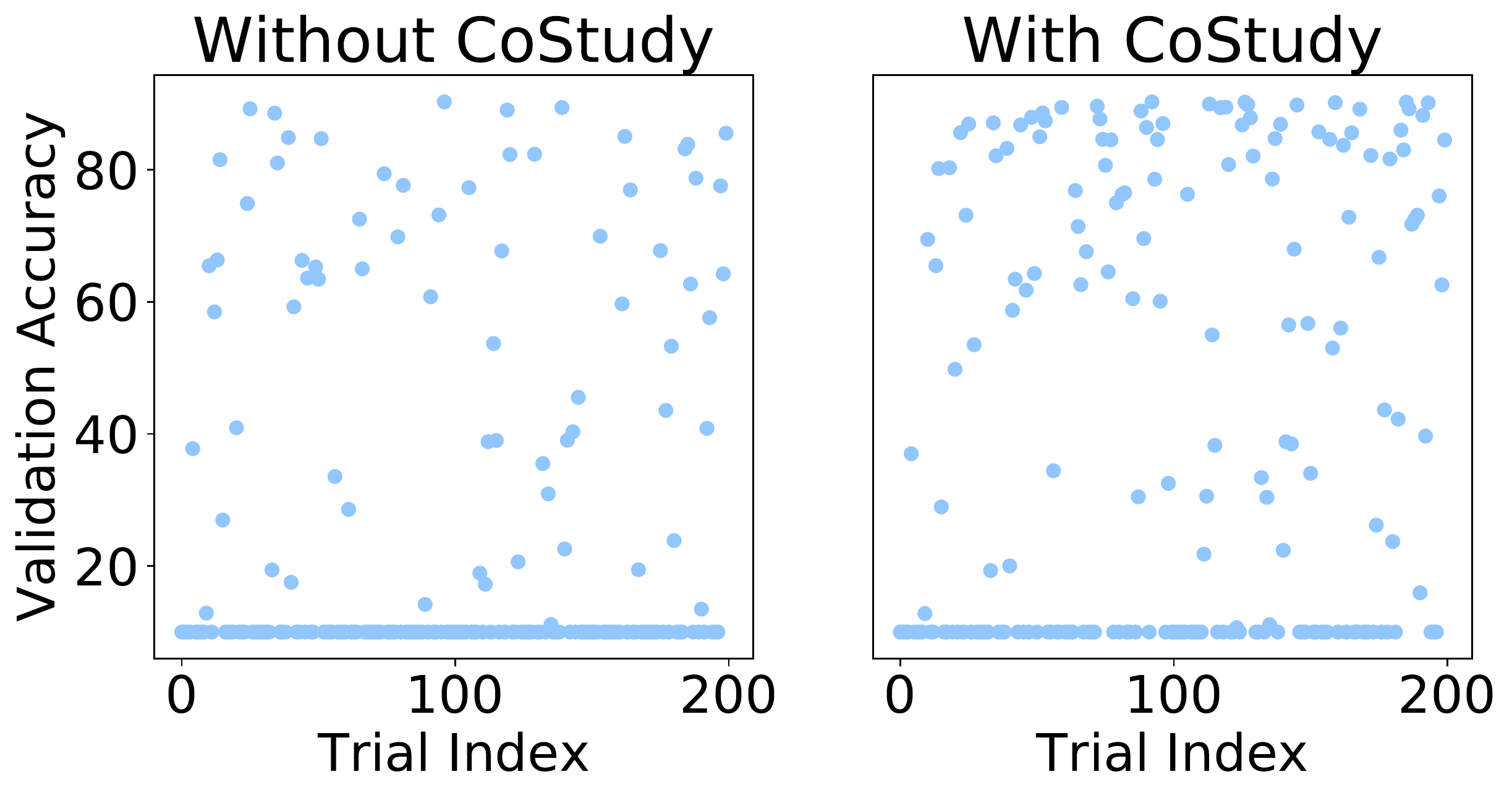}
		\caption{\label{fig:rnd-scatter}}
	\end{subfigure}
	\begin{subfigure}{0.25\textwidth}
		\includegraphics[width=0.9\textwidth]{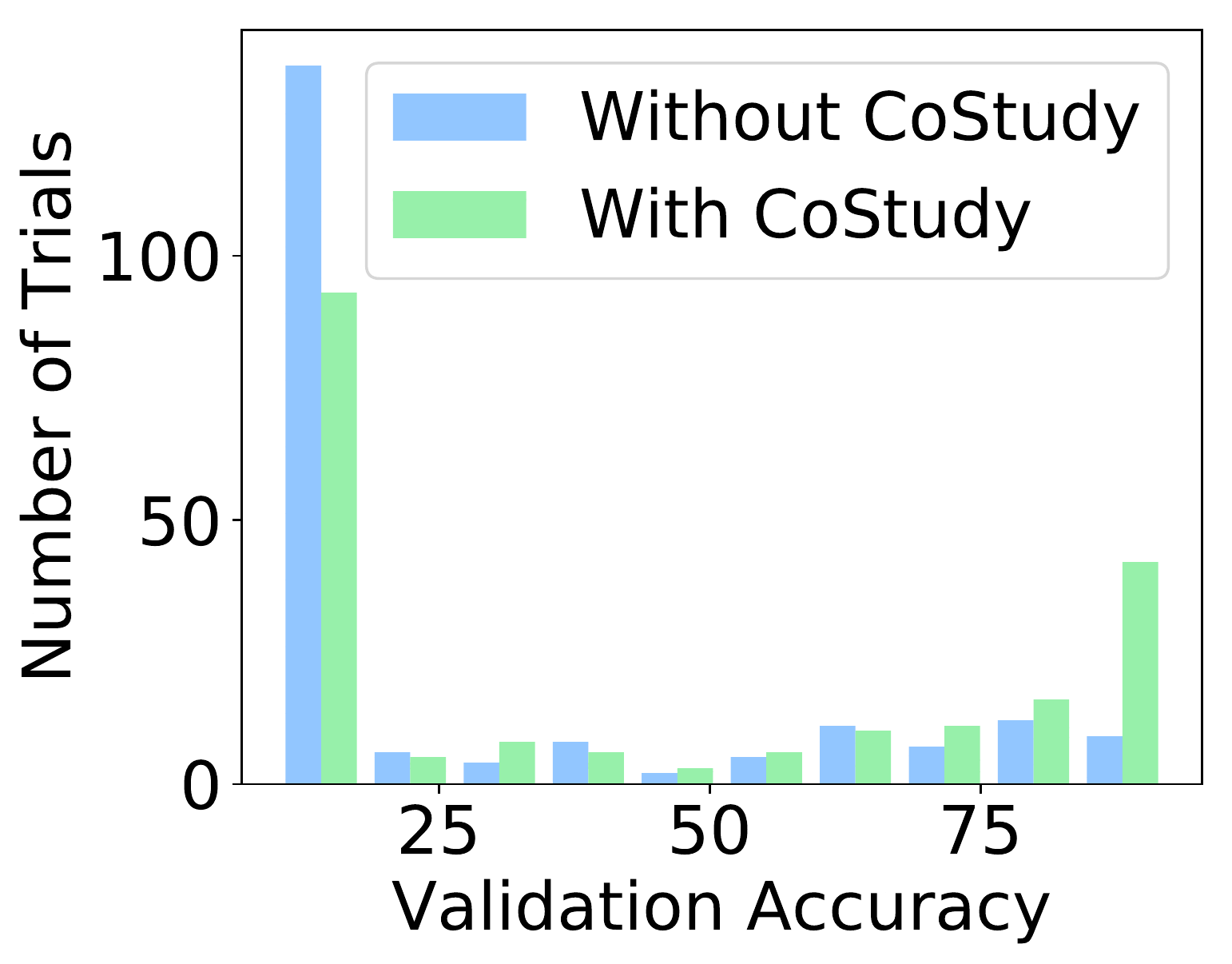}
		\caption{\label{fig:rnd-hist}}
	\end{subfigure}
	\begin{subfigure}{0.25\textwidth}
		\includegraphics[width=0.9\textwidth]{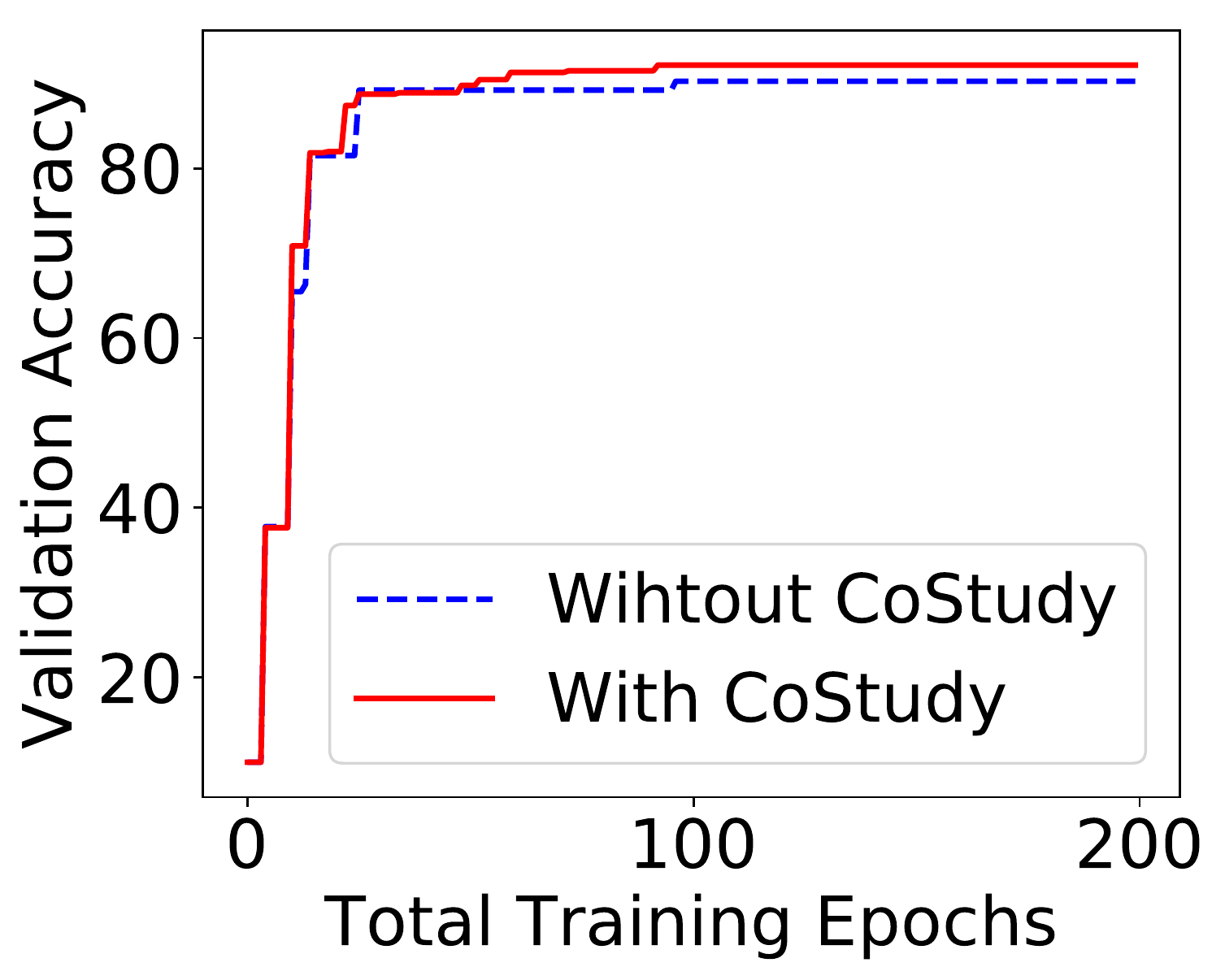}
		\caption{\label{fig:rnd-line}}
	\end{subfigure}
	\caption{Hyper-parameter tuning based on random search.\label{fig:random}}
\end{figure*}

\begin{figure*}[h]
	\centering
	\begin{subfigure}{0.43\textwidth}
		\centering
		\includegraphics[width=0.9\textwidth]{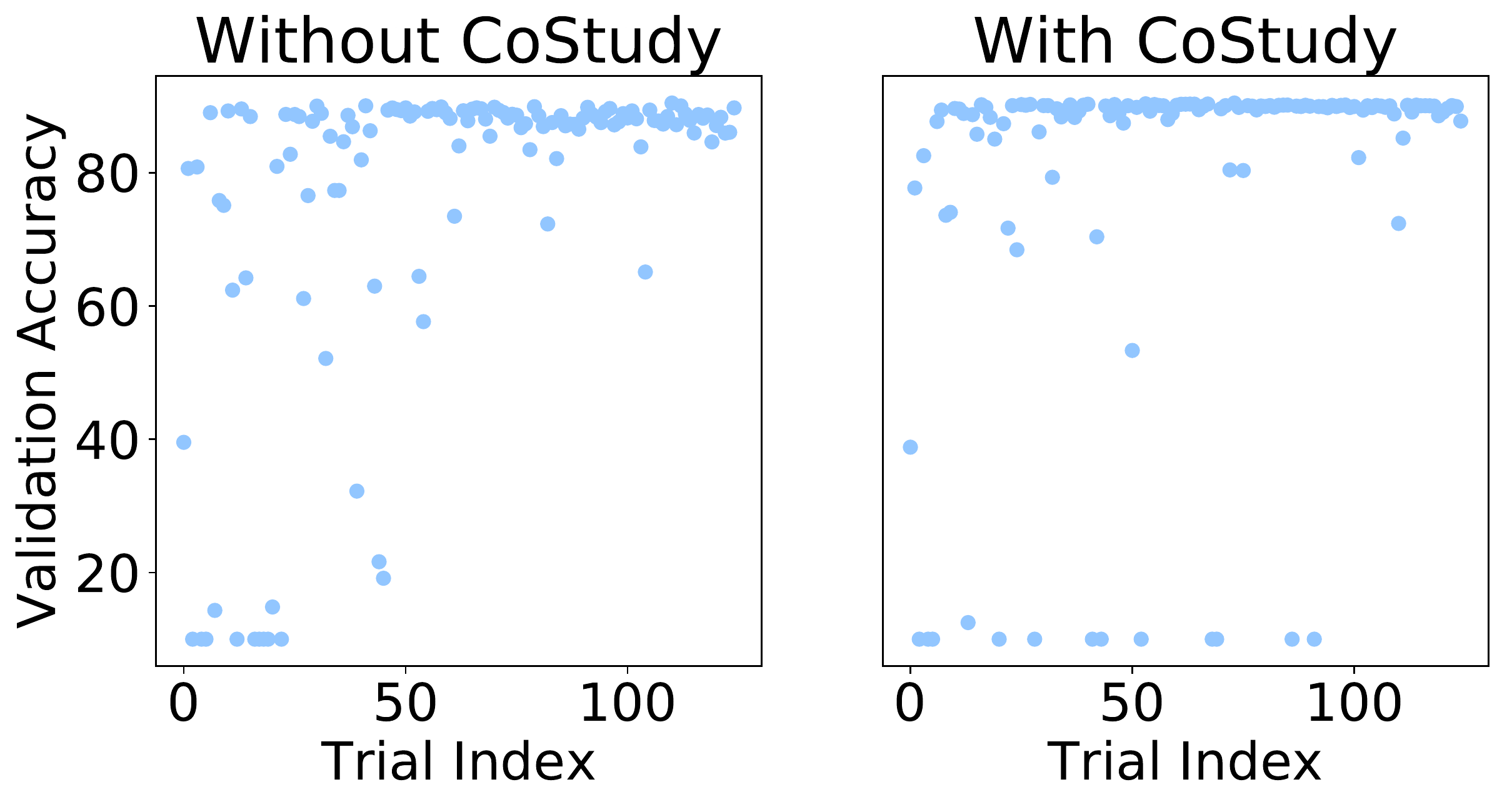}
		\caption{\label{fig:bo-scatter}}
	\end{subfigure}
	\begin{subfigure}{0.25\textwidth}
		\includegraphics[width=0.9\textwidth]{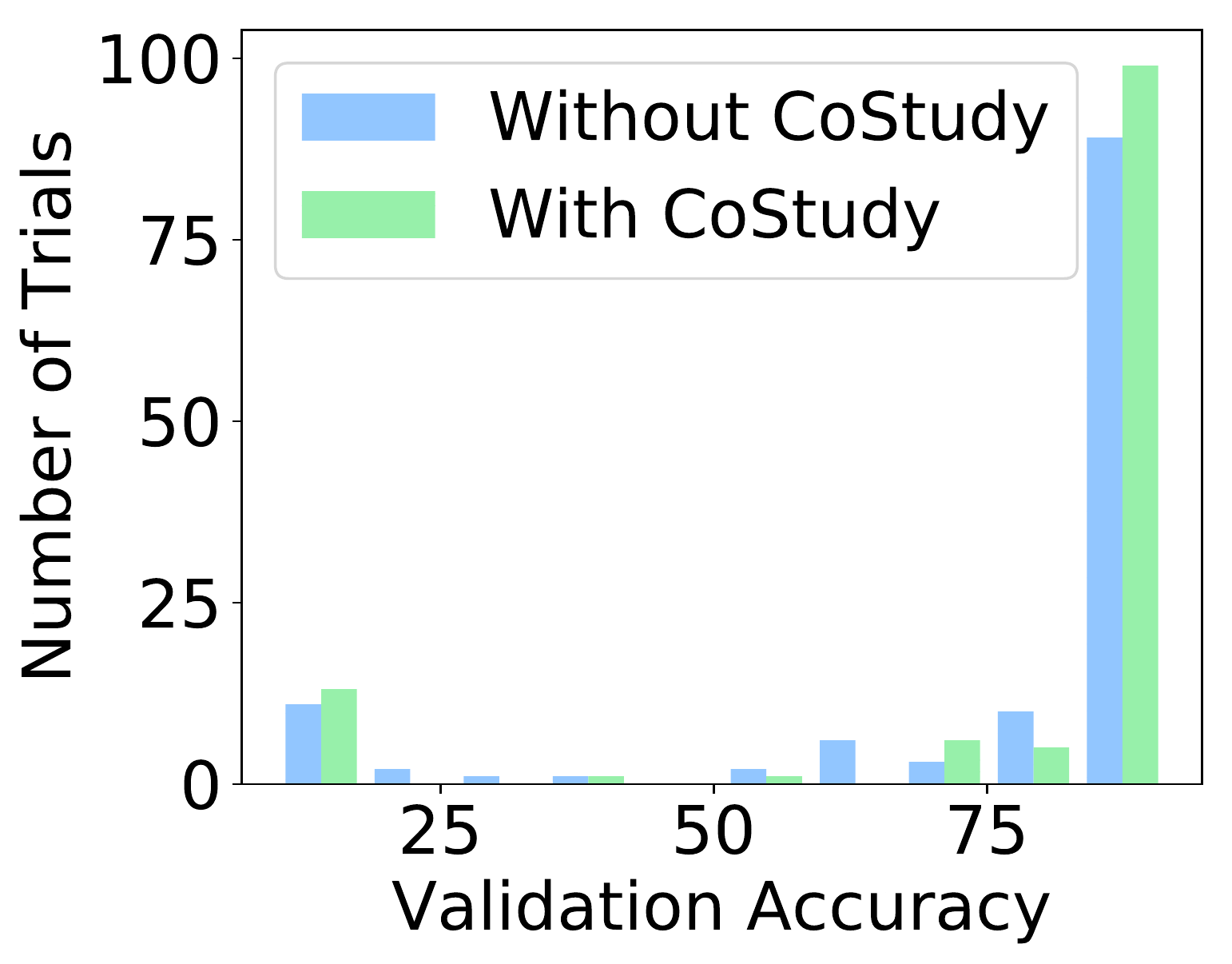}
		\caption{\label{fig:bo-hist}}
	\end{subfigure}
	\begin{subfigure}{0.25\textwidth}
		\includegraphics[width=0.9\textwidth]{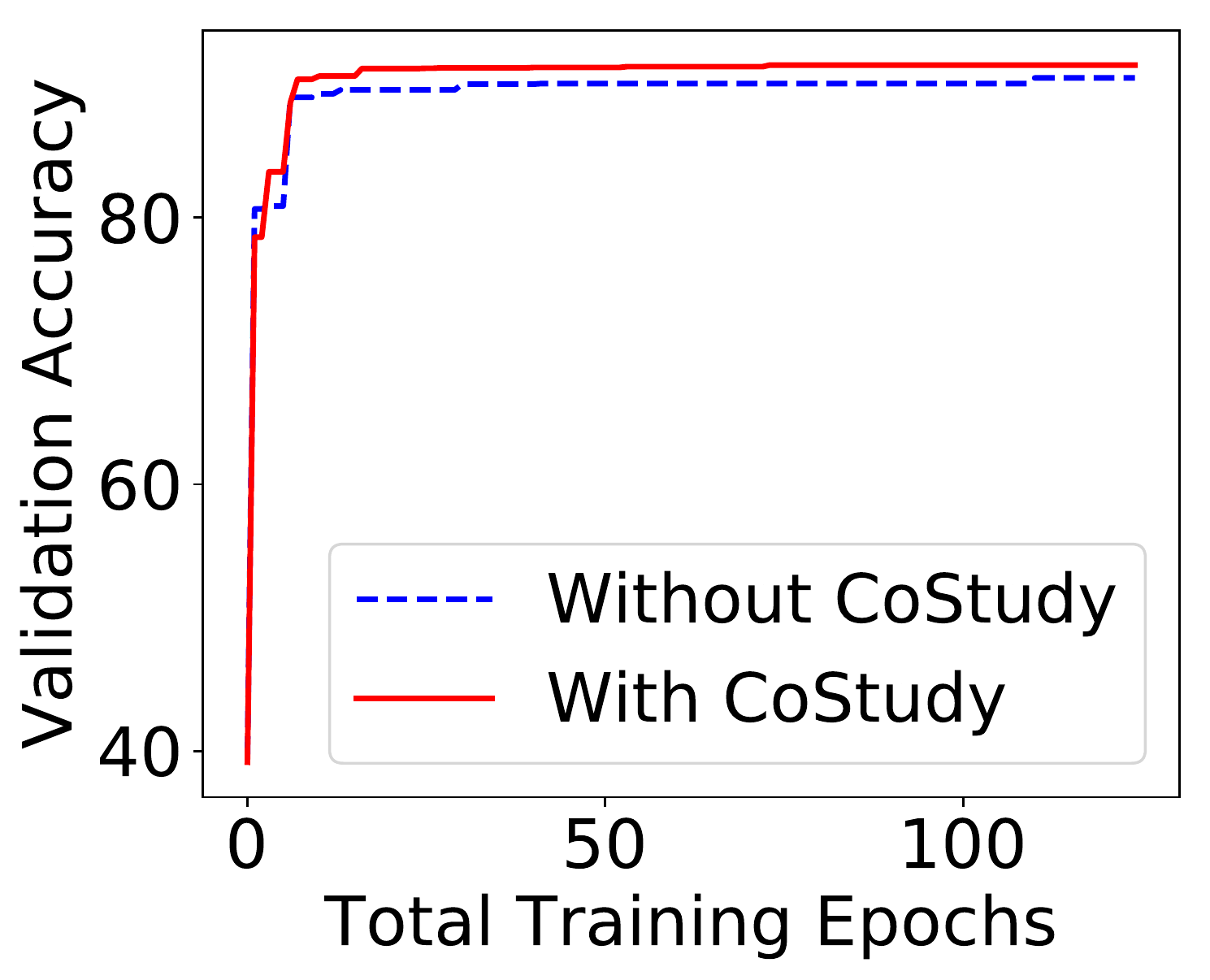}
		\caption{\label{fig:bo-line}}
	\end{subfigure}
	\caption{Hyper-parameter tuning based on Bayesian optimization.\label{fig:bo}}
\end{figure*}

\subsection{Failure Recovery}

For both the hyper-parameter training and inference services, the workers are stateless. 
Hence, Rafiki manager can easily recover these nodes by running a new docker container and registering it into the training or inference master. 
However, for the masters, they have state information. For example, the master for the training service records the current best hyper-parameter trial. 
The master for the inference service has the state, action and reward for reinforcement learning.
Rafiki checkpoints these (small) state information of masters for fast failure recovery.

\section{Experimental Study}\label{sec:exp}
In this section we evaluate the scalability, efficiency and effectiveness of Rafiki for hyper-parameter tuning and inference service optimization respectively. 
The experiments are conducted on three machines, each with 3 Nvidia GTX 1080Ti GPUs,  1 Intel E5-1650v4 CPU and 64GB memory.

\subsection{Evaluation of Hyper-parameter Tuning}

\textbf{Task and Dataset} We test Rafiki's distributed hyper-parameter tuning service by running it to tune the hyper-parameters of deep ConvNets over CIFAR-10 for image classification. 
CIFAR-10 is a popular benchmark dataset with RGB images from 10 categories. 
Each category has 5000 training images (including 1000 validation images) and 1000 test images. 
All images are of size 32x32. 
There is a standard sequence of preprocessing steps for CIFAR-10, which subtracts the mean and divides the standard deviation from each channel computed on the training images, pads each image with 4 pixels of zeros on each side to obtain a 40x40 pixel image, randomly crops a 32x32 patch, and then flips (horizontal direction) the image randomly with probability 0.5.

\subsubsection{Tuning Optimization Hyper-parameters}
We fix the ConvNet architecture to be the same as shown in Table 5 of \cite{2015arXiv150205700S}, which has 8 convolution layers. The hyper-parameters to be tuned are from the optimization algorithm, including momentum, learning rate, weight decay coefficients, dropout rates, and standard deviations of the Gaussian distribution for weight initialization. 
We run each trial with early stopping, which terminates the training when the validation loss stops decreasing.

We compare the na\"ive distributed tuning algorithm, i.e. \emph{Study} (Algorithm~\ref{alg:master}) and the collaborative tuning algorithm. i.e. \emph{CoStudy} (Algorithm~\ref{alg:cmaster}) with different \emph{TrialAdvisor} algorithms.  
Figure~\ref{fig:random} shows the comparison using random search~\cite{Bergstra:2012:RSH:2188385.2188395} for \emph{TrialAdvisor}. 
In particular, each point in Figure~\ref{fig:rnd-scatter} stands for one trial. 
We can see that the top area for CoStudy is denser than that for Study. 
In other words, CoStudy is more likely to  get better performance.
This is confirmed in Figure~\ref{fig:rnd-hist}, which shows that CoStudy has more trials with high accuracy (i.e. accuracy $>$50\%) than Study, and has fewer trials with low accuracy (i.e., accuracy $\leq$50\%). Figure~\ref{fig:rnd-line} illustrates the tuning progress of the two approaches, where each point on the line represents the best performance among all trials conducted so far. 
The x-axis is the total number of training epochs\footnote{training the model by scanning the dataset once is called one epoch.}, which corresponds to the tuning time (=total number of epochs times the time per epoch). We can observe that CoStudy is faster than Study and achieves better accuracy than Study. 
Notice that the validation accuracy is very high ($>$91\%). Therefore, a small difference (1\%) indicates significant improvement.

\begin{figure*}[h!]
	\centering
	\includegraphics[width=0.8\textwidth]{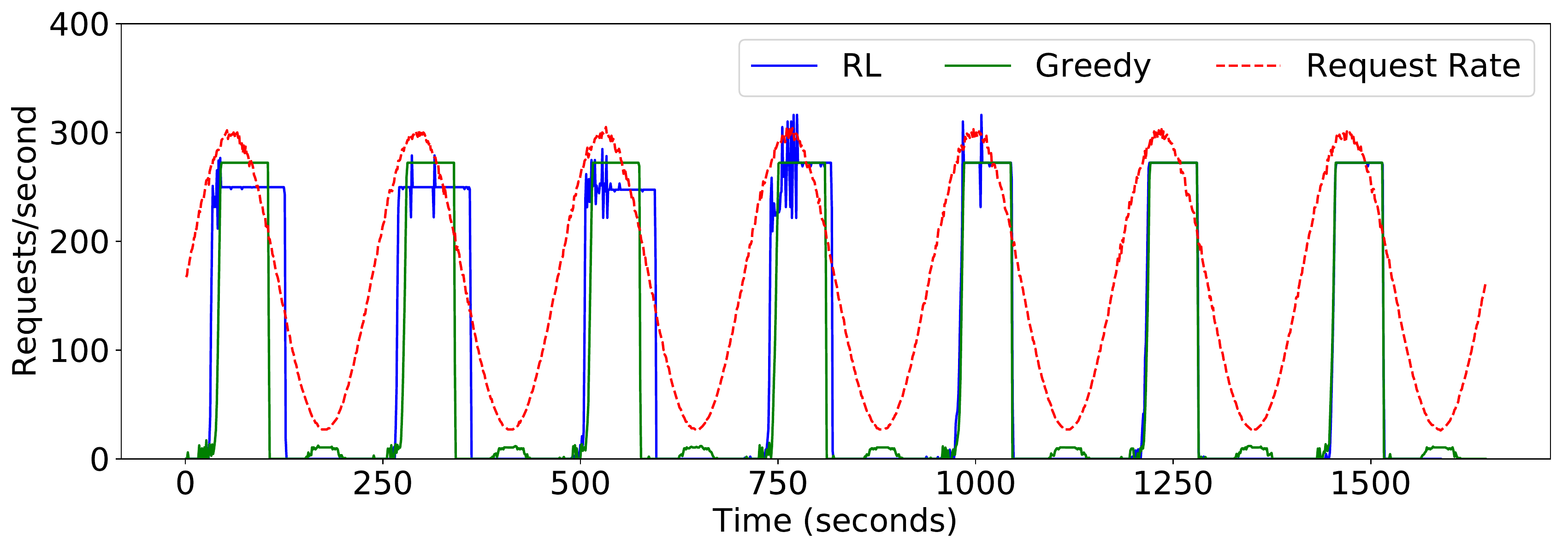}
	\caption{Single inference model with the arriving rate defined based on the maximum throughput. \label{fig:single-max}}
\end{figure*}

\begin{figure}[h!]
	\centering
	\begin{subfigure}{0.235\textwidth}
		\centering
		\includegraphics[width=0.9\textwidth]{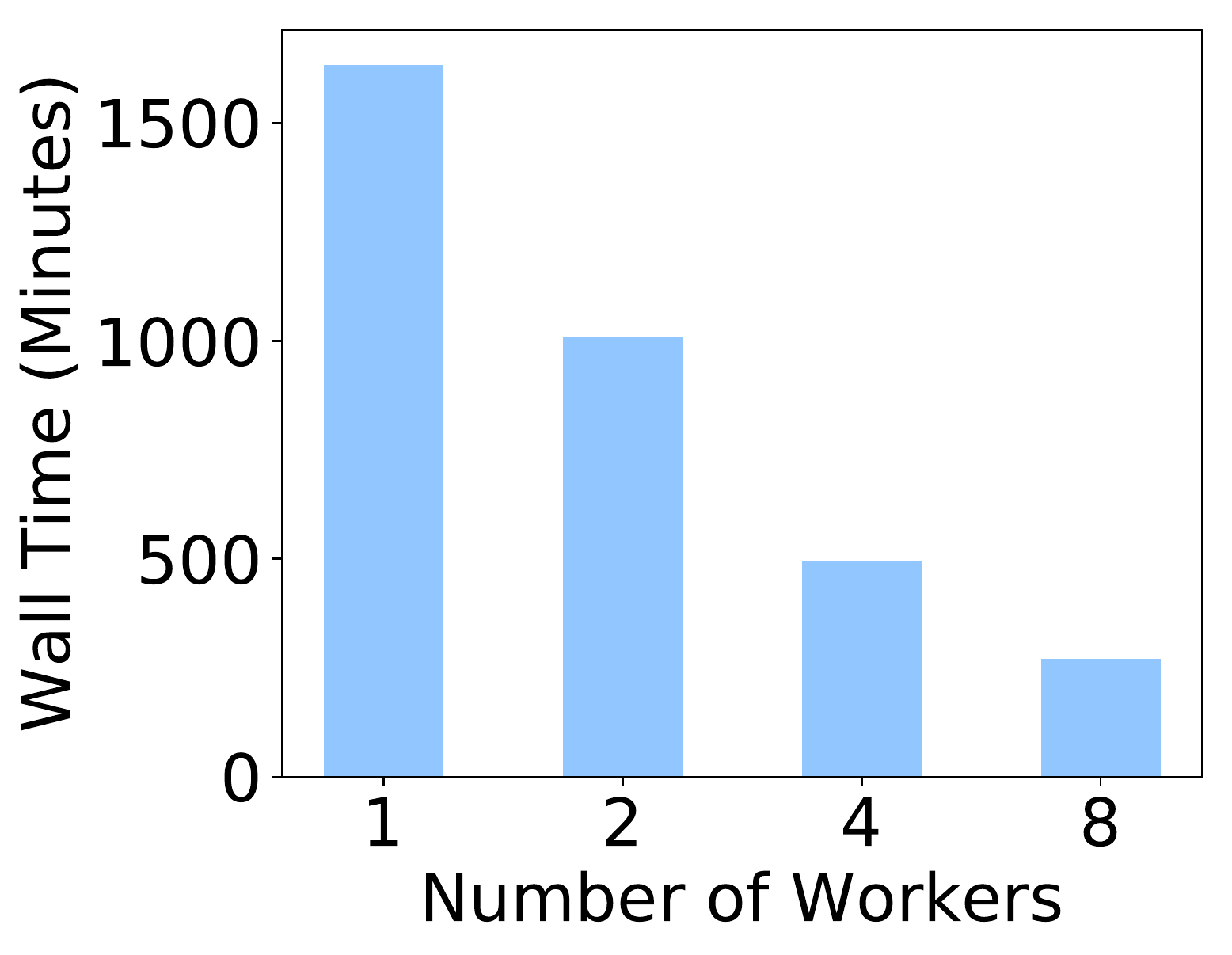}
		\caption{\label{fig:scale-bar}}
	\end{subfigure}
	\begin{subfigure}{0.235\textwidth}
		\includegraphics[width=0.9\textwidth]{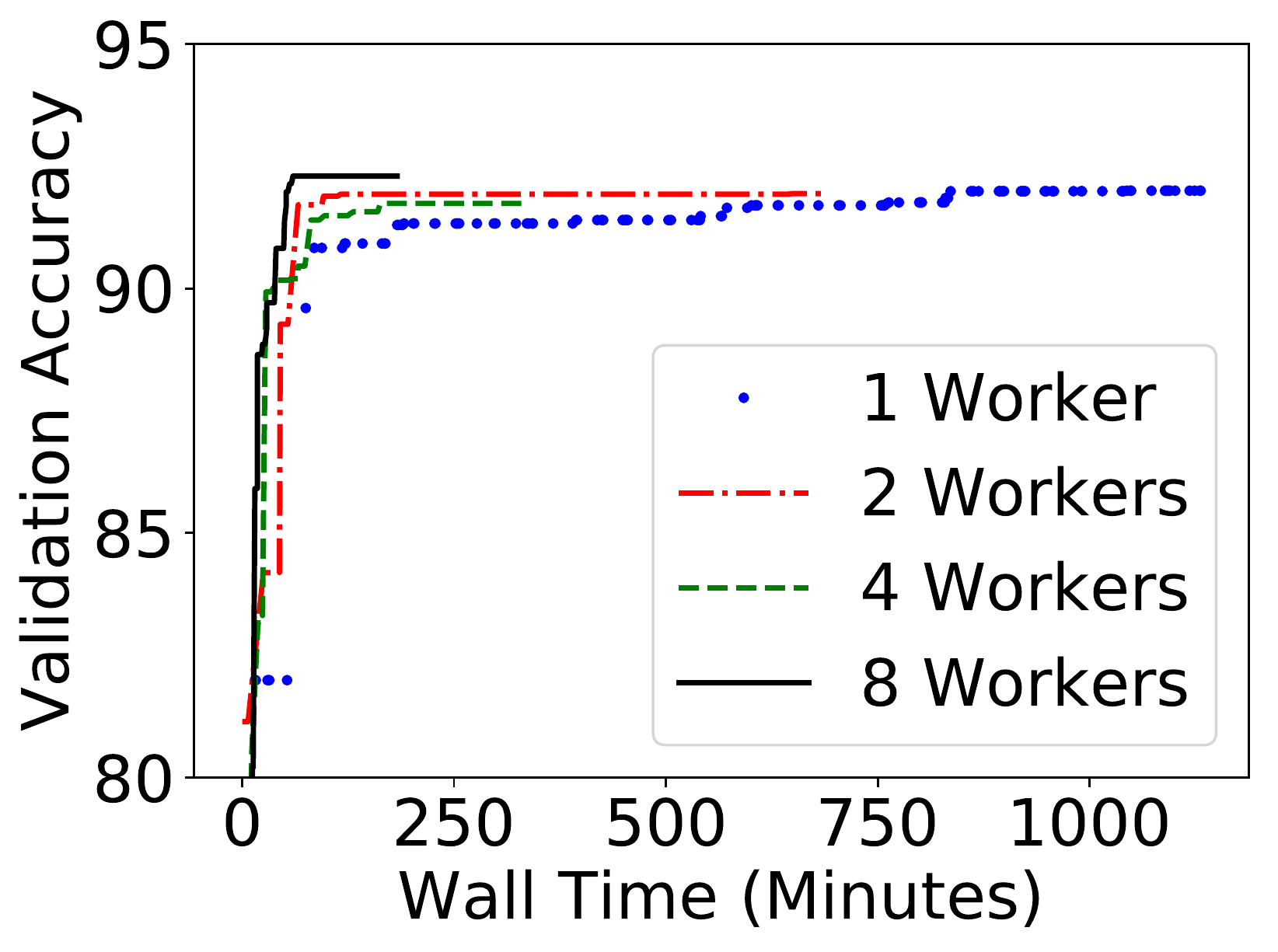}
		\caption{\label{fig:scale-line}}
	\end{subfigure}
	\caption{Scalability test of distributed hyper-parameter tuning.\label{fig:scalability}}
\end{figure}

Figure~\ref{fig:bo} compares Study and CoStudy using Gaussian process based Bayesian Optimization (BO)\footnote{https://github.com/scikit-optimize} for TrialAdvisor. 
Comparing Figure~\ref{fig:bo-scatter} and Figure~\ref{fig:rnd-scatter}, we can see there are more points in the top area of BO figures. 
In other words, BO is better than random search，which has been observed in other papers~\cite{2012arXiv1206.2944S}. In Figure~\ref{fig:bo-scatter}, CoStudy has a few more points in the right bottom area than CoStudy. 
After doing an in-depth inspection, we found that those points were trials initialized randomly instead of from pre-trained models. For Study, it always uses random initialization, hence the BO algorithm has a fixed prior about the initialization method. However, for CoStuy, its initialization is from pre-trained models for most time. These random initialization trials change the prior and thus get biased estimation about the Gaussian process, which leads to poor accuracy. Since we are decaying $\alpha$ to reduce the chance of random initialization, there are fewer and fewer points in the right bottom area. Overall, CoStudy achieves better accuracy as shown in Figure~\ref{fig:bo}b and Figure~\ref{fig:bo}c.

\begin{figure}[h!]
	\centering
	\includegraphics[width=.4\textwidth]{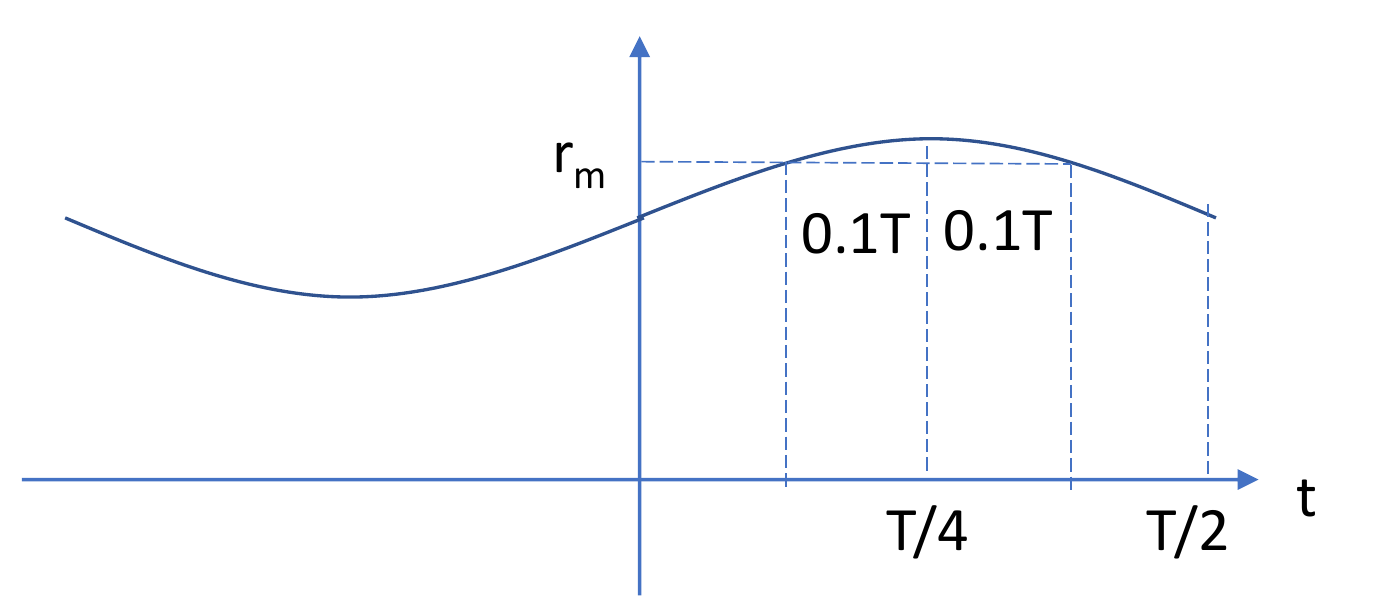}
	\caption{Sine function for controlling the request arriving rate.\label{fig:sine}}
\end{figure}

We study the scalability of Rafiki by varying the number of workers. Figure~\ref{fig:scalability} compares the 3 jobs running over 1, 2, 4 and 8 GPUs respectively. 
Each point on the curves in the figure represents the best validation performance among all trials that have been tested so far. The x-axis is the wall clock time. 
We can see that with more GPUs, the tuning becomes faster. It scales almost linearly. 

\begin{figure*}[h!]
	\centering
	\includegraphics[width=0.7\textwidth]{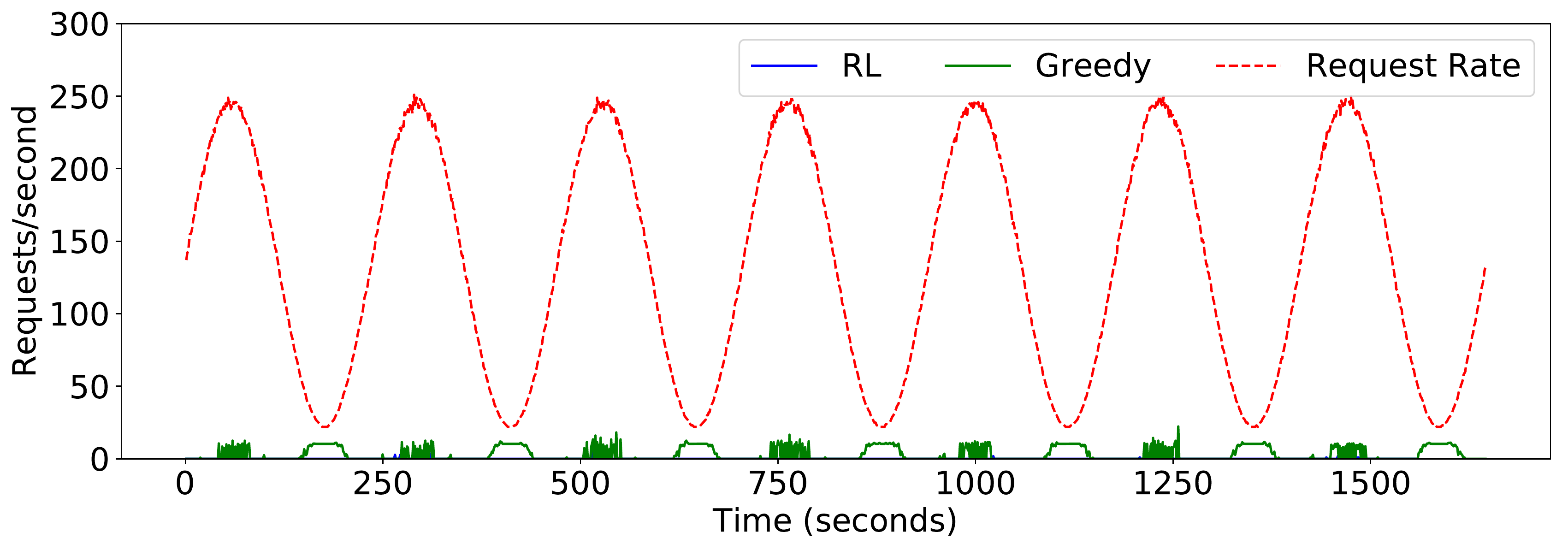}
	\caption{Single inference model with the arriving rate defined based on the minimum throughput. \label{fig:single-min}}
\end{figure*}

\subsection{Evaluation of Inference Optimization}
We use image classification as the application to test the optimization techniques introduced in Section~\ref{sec:infer}. 
The inference models are ConvNets trained over the ImageNet~\cite{DBLP:conf/nips/KrizhevskySH12} benchmark dataset. ImageNet is a popular image classification benchmark with many open-source ConvNets trained on it (Figure~\ref{fig:convnets}). It has 1.2 million RGB training images and 50,000 validation images.

\begin{figure*}[h!]
	\centering
	\begin{subfigure}{0.255\textwidth}
		\centering
		\includegraphics[width=\textwidth]{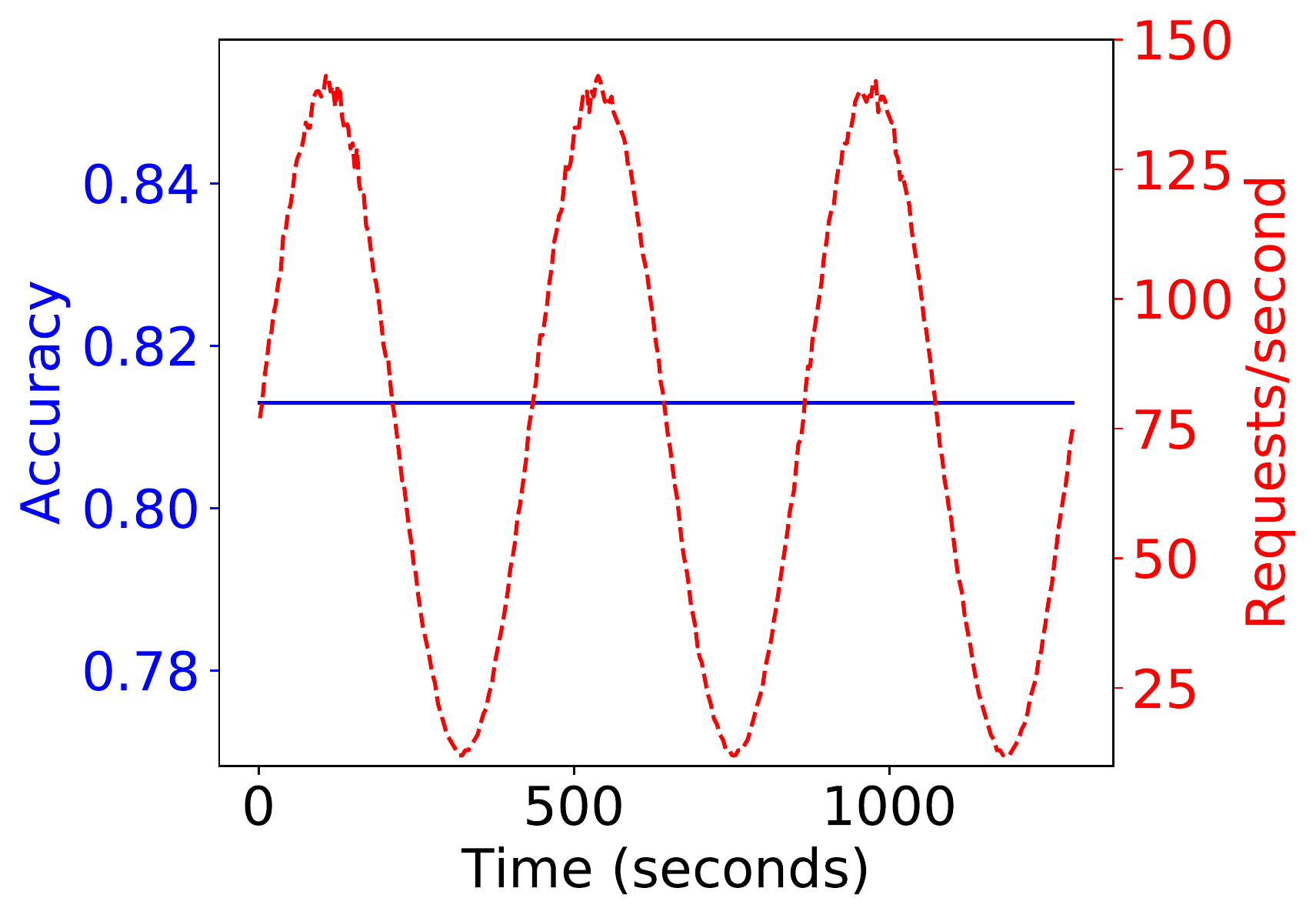}
		\caption{Greedy algorithm. \label{fig:acc-sync}}
	\end{subfigure}
	\begin{subfigure}{0.27\textwidth}
		\centering
		\includegraphics[width=\textwidth]{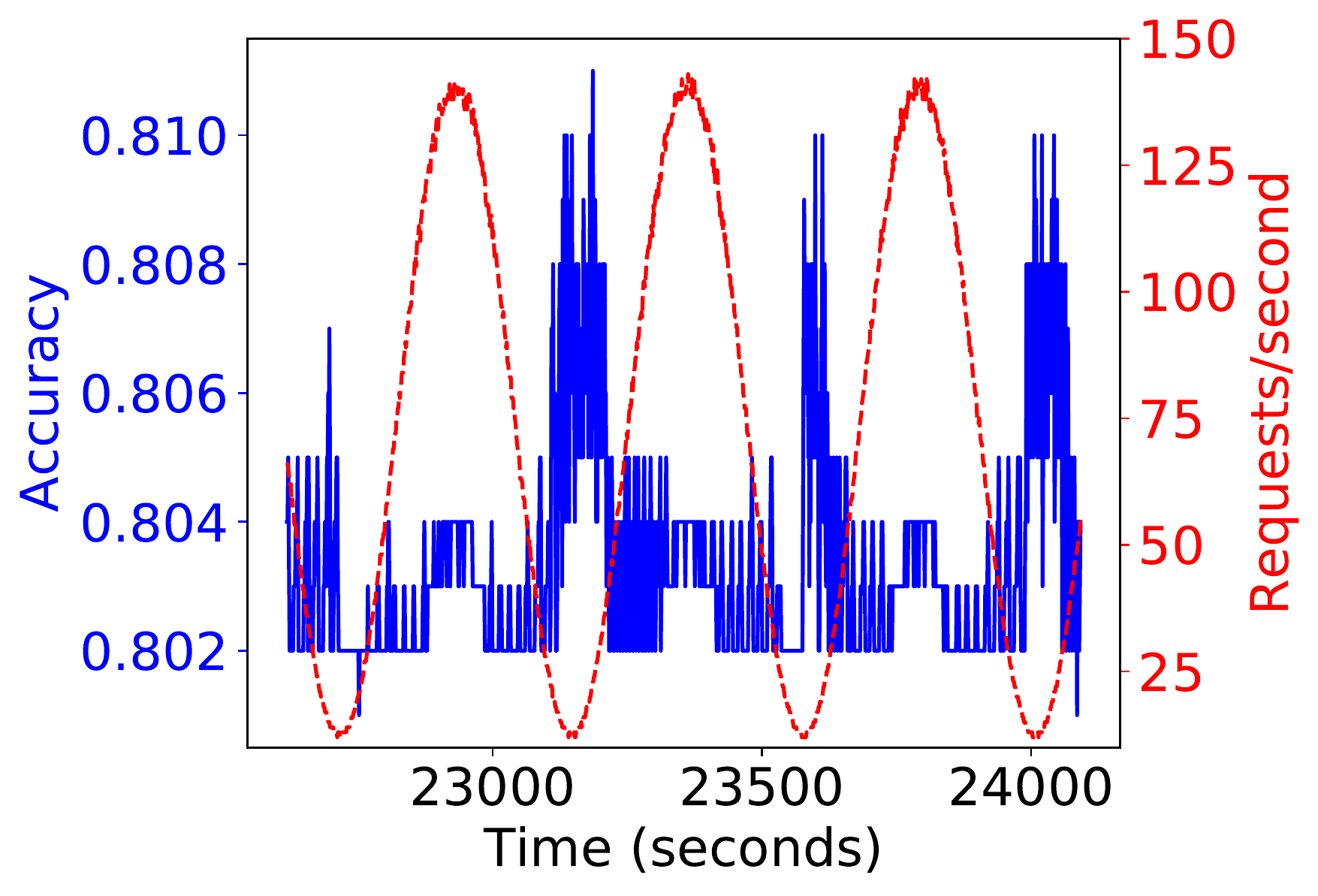}
		\caption{RL algorithm. \label{fig:acc-sync1}}
	\end{subfigure}
	\begin{subfigure}{0.23\textwidth}
		\includegraphics[width=\textwidth]{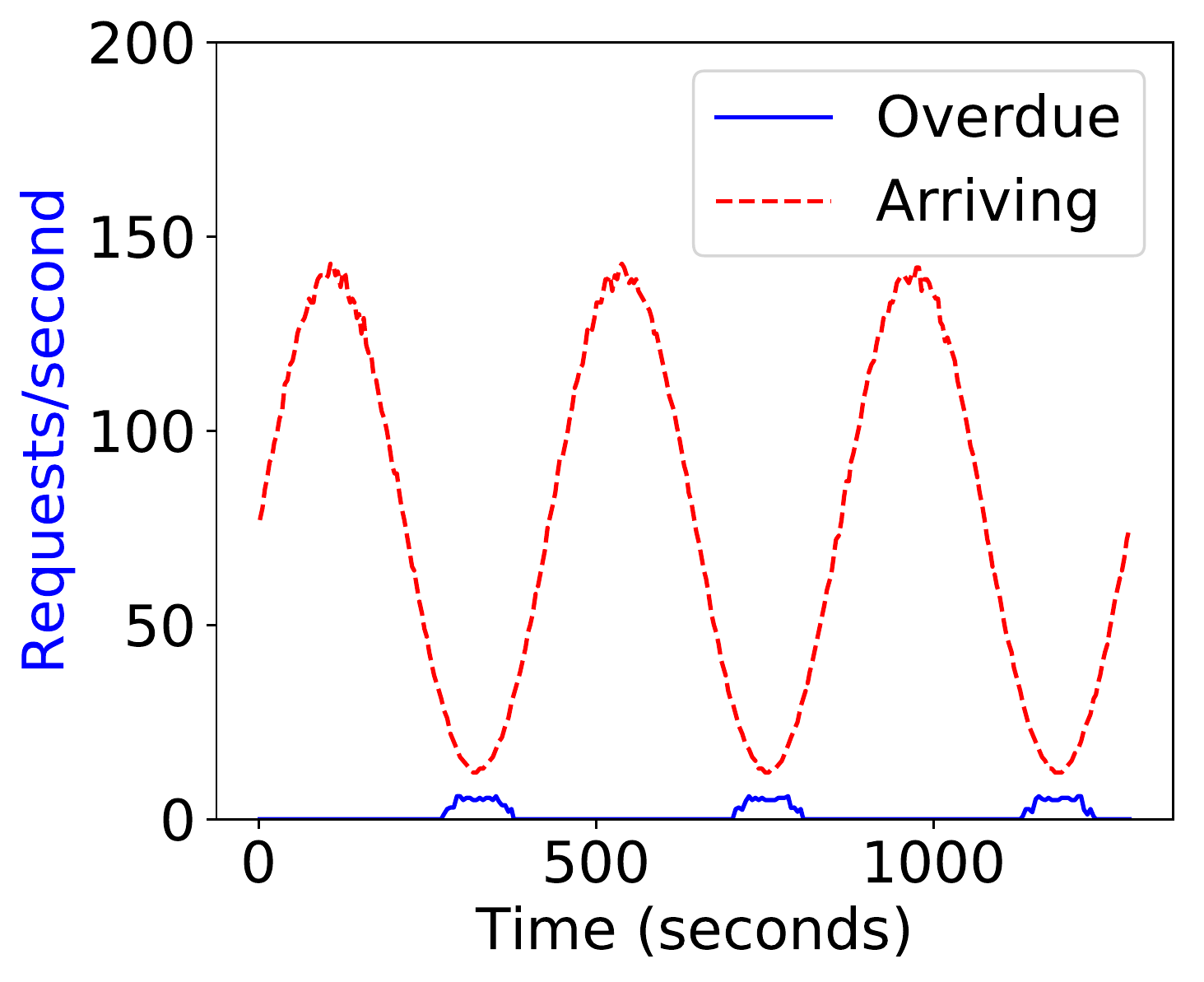}
		\caption{Greedy algorithm. \label{fig:overdue-sync}}
	\end{subfigure}
	\begin{subfigure}{0.23\textwidth}
		\includegraphics[width=\textwidth]{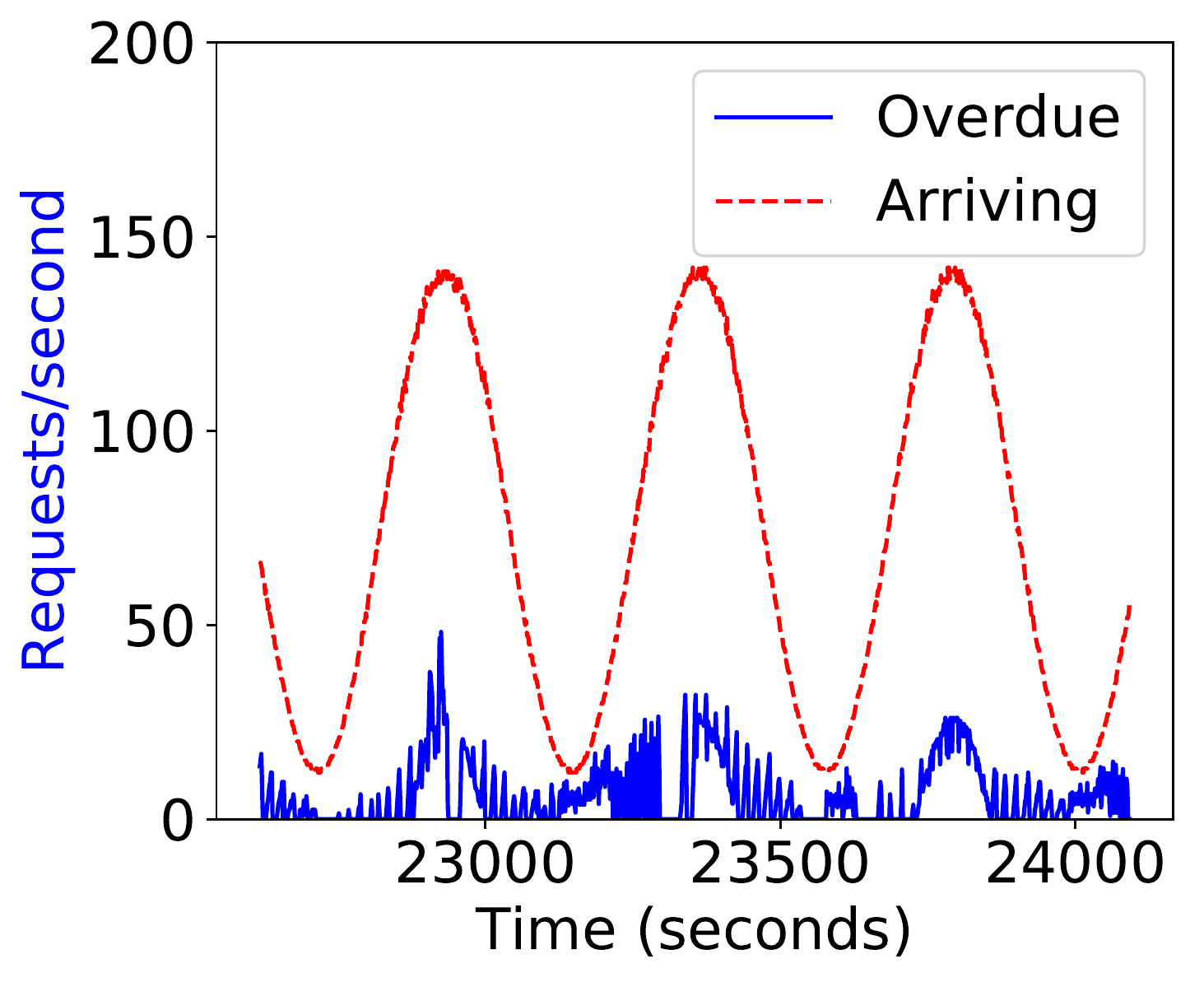}
		\caption{RL algorithm. \label{fig:overdue-sync1}}
	\end{subfigure}
	\caption{Multiple model inference with the arriving rate defined based on the minimum throughput.\label{fig:sync}}
\end{figure*}

\begin{figure*}[h!]
	\centering
	\begin{subfigure}{0.255\textwidth}
		\centering
		\includegraphics[width=\textwidth]{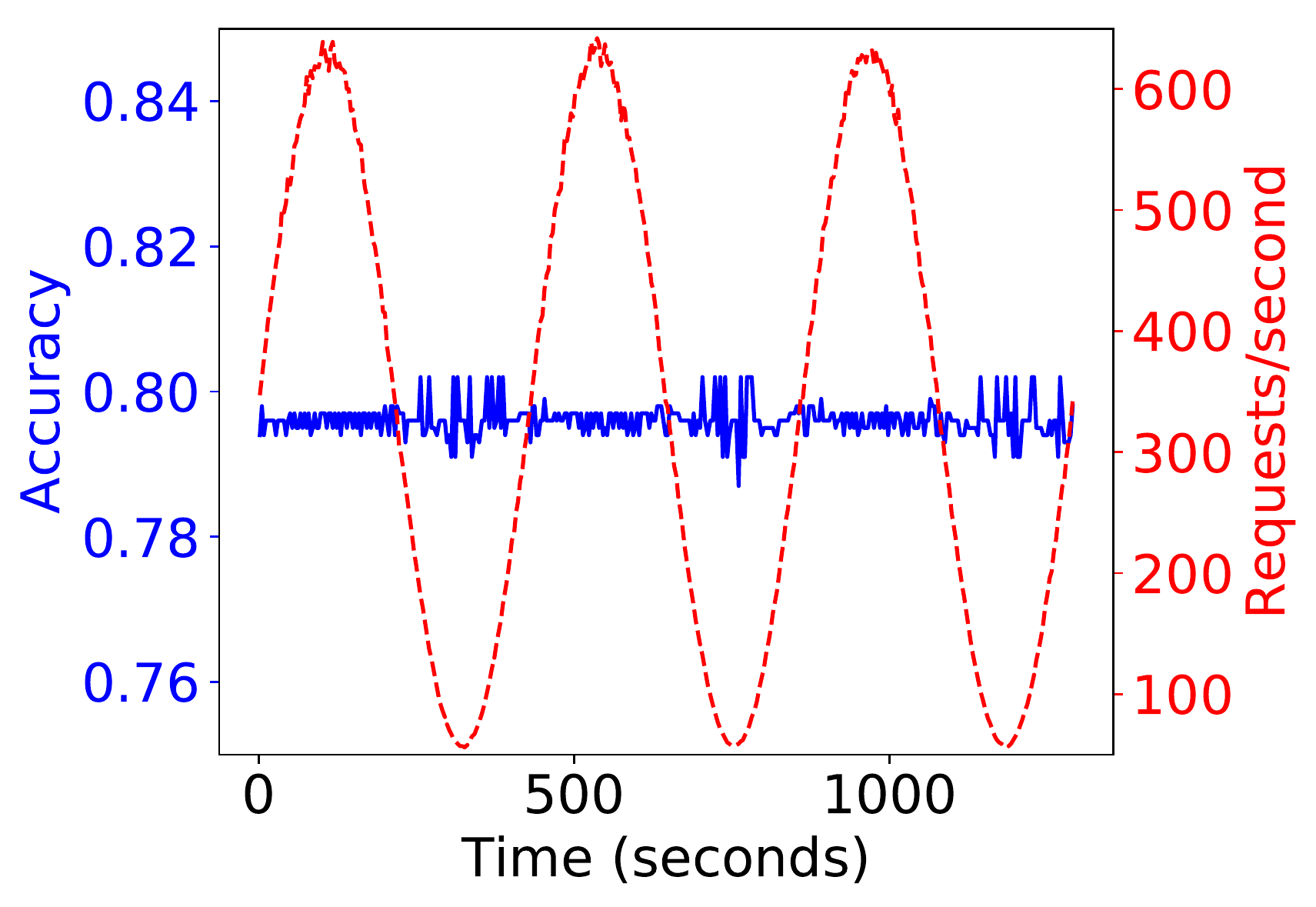}
		\caption{Greedy algorithm. \label{fig:acc-async}}
	\end{subfigure}
	\begin{subfigure}{0.255\textwidth}
		\centering
		\includegraphics[width=\textwidth]{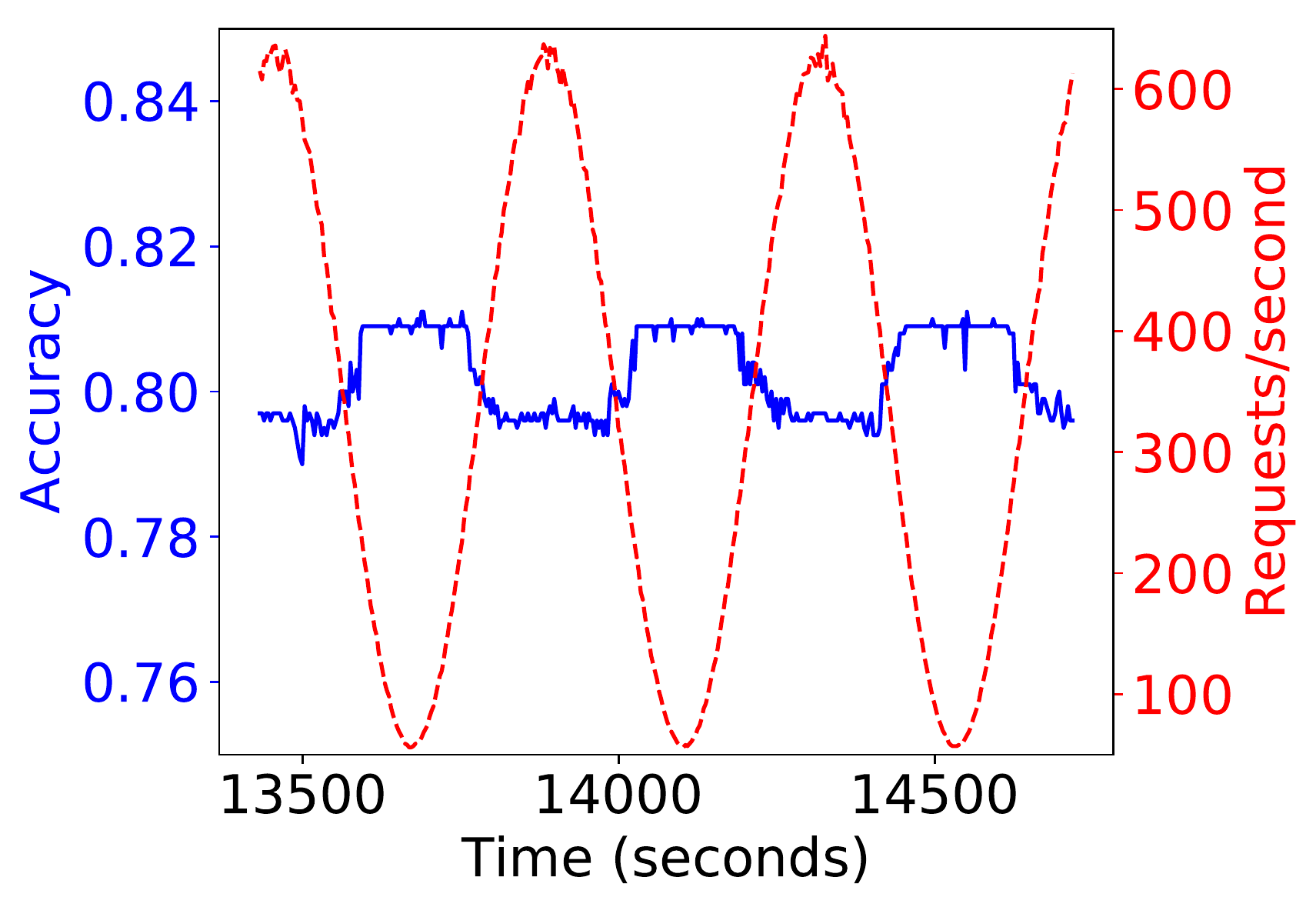}
		\caption{RL algorithm. \label{fig:acc-async1}}
	\end{subfigure}
	\begin{subfigure}{0.23\textwidth}
		\includegraphics[width=\textwidth]{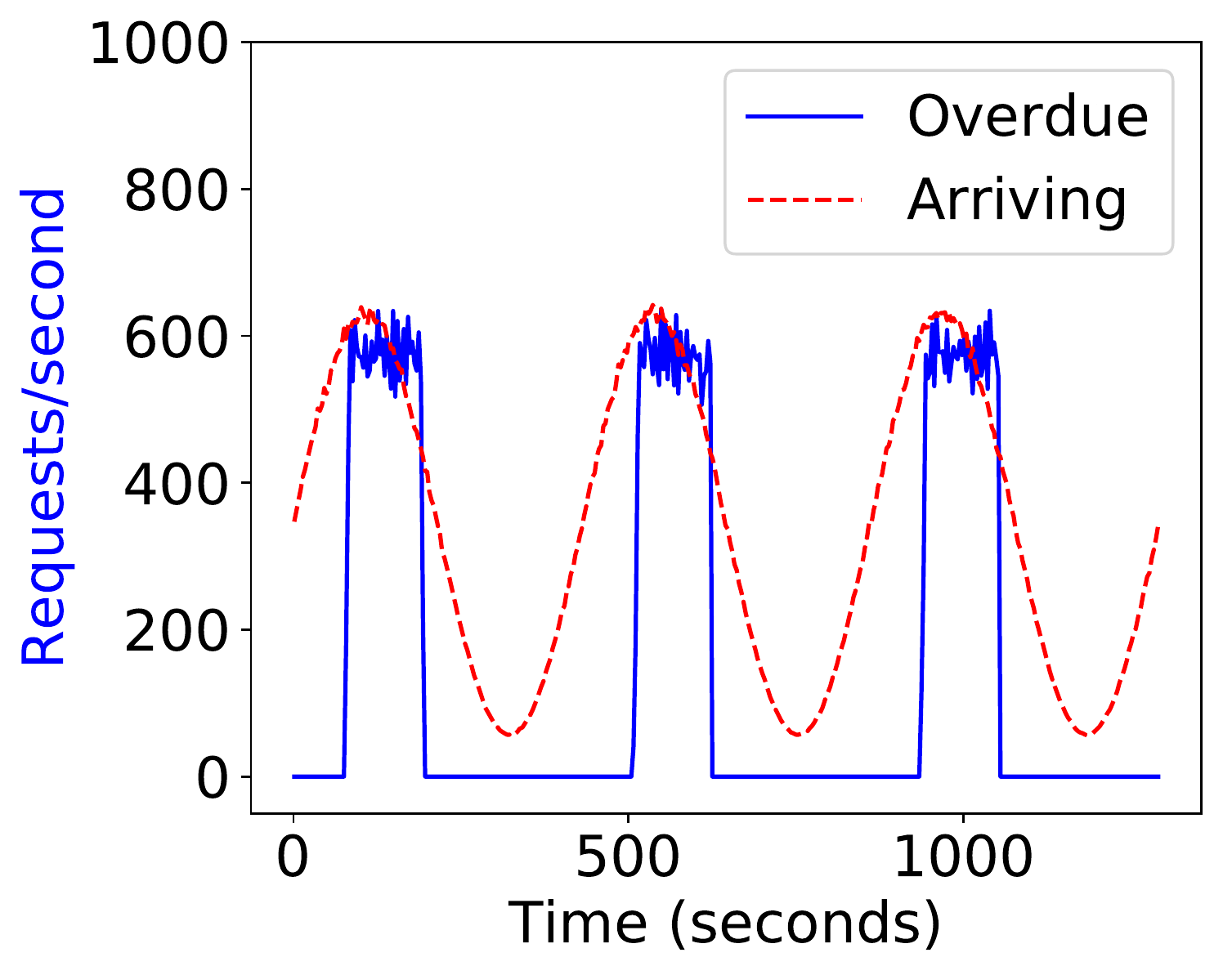}
		\caption{Greedy algorithm. \label{fig:overdue-async}}
	\end{subfigure}
	\begin{subfigure}{0.23\textwidth}
		\includegraphics[width=\textwidth]{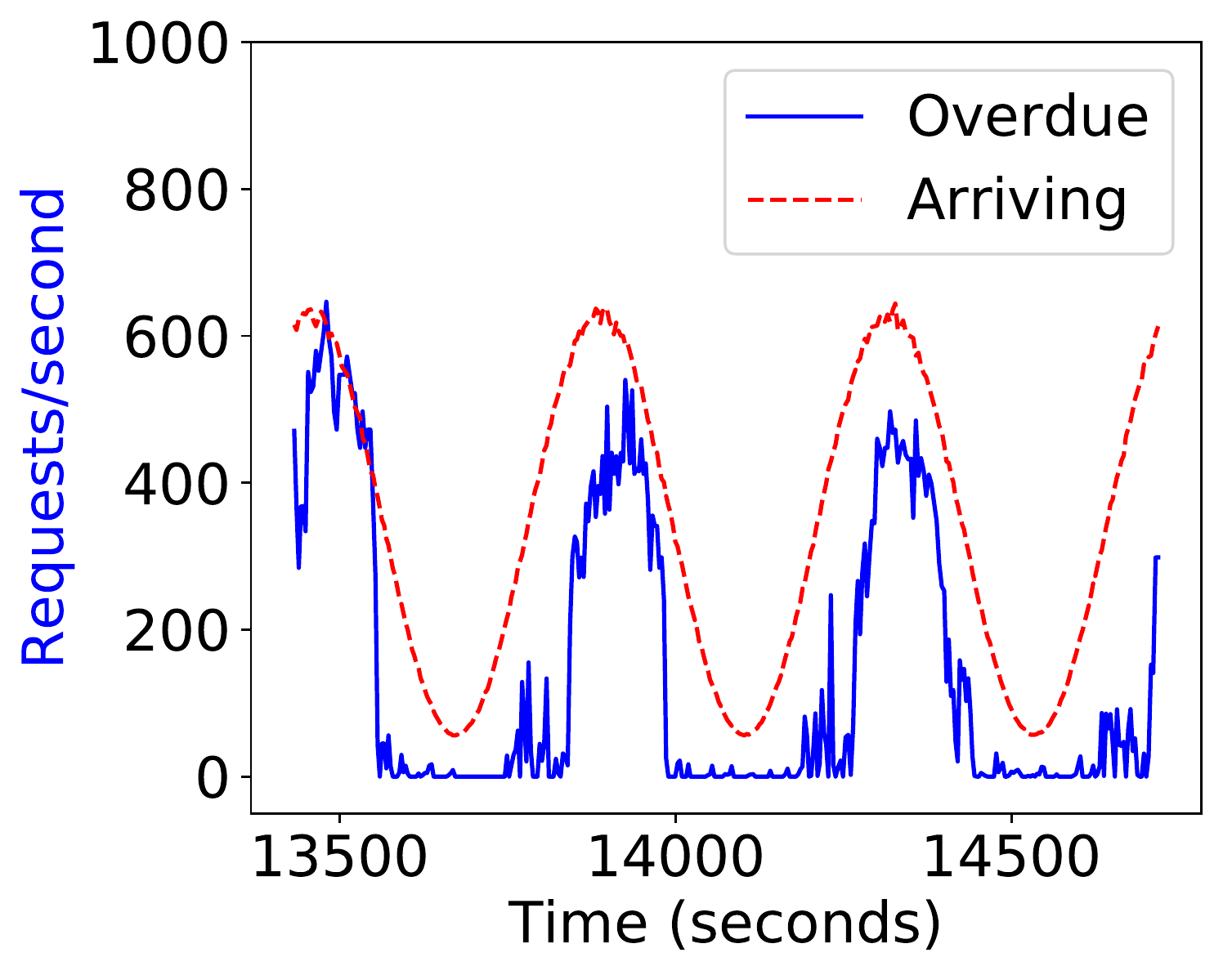}
		\caption{RL algorithm. \label{fig:overdue-async1}}
	\end{subfigure}
	\caption{Multiple model inference with the arriving rate defined based on the maximum throughput.\label{fig:async}}
\end{figure*}

\begin{figure*}[h!]
	\centering
	\begin{subfigure}{0.255\textwidth}
		\centering
		\includegraphics[width=\textwidth]{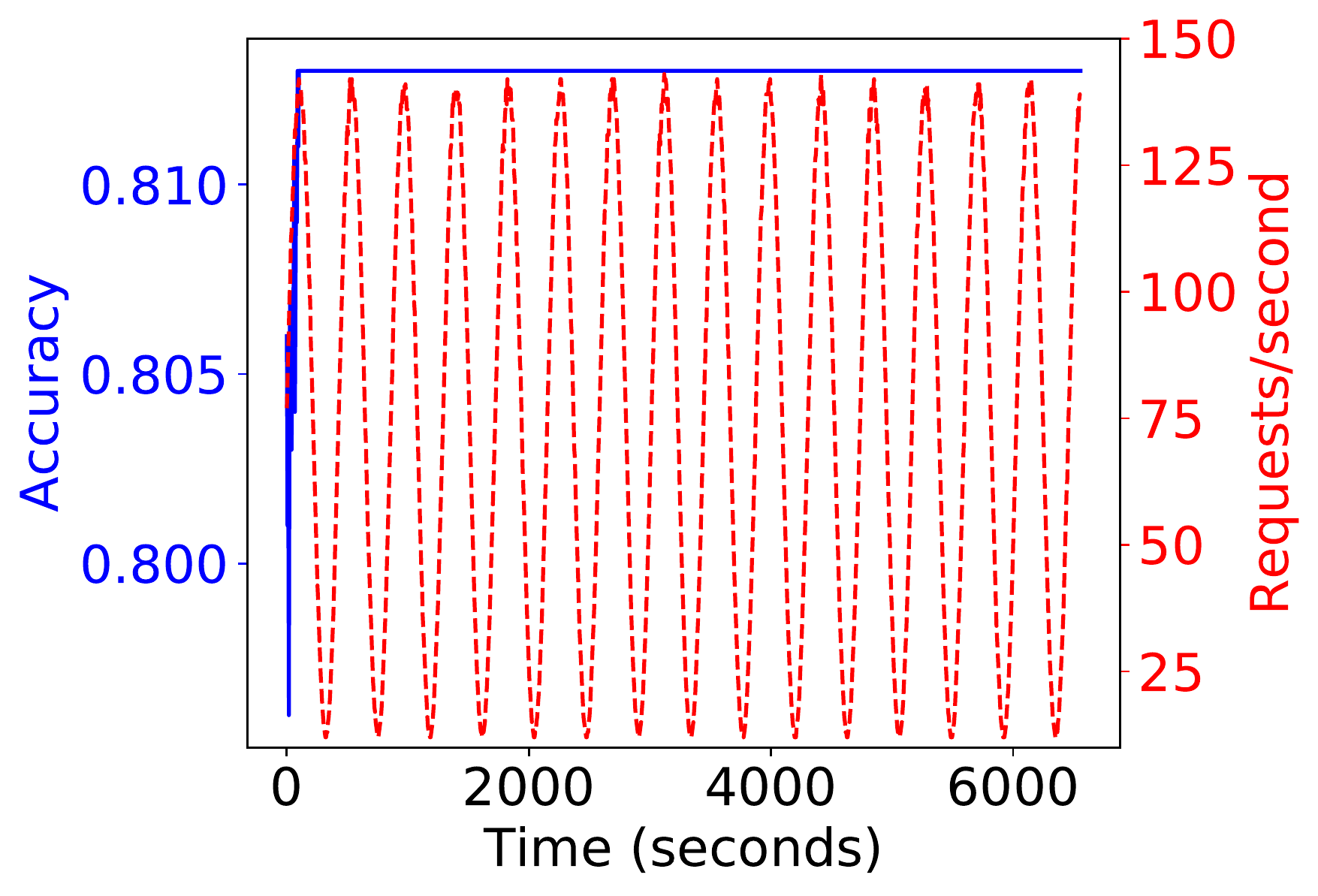}
		\caption{$\beta=0$. \label{fig:acc-sync0}}
	\end{subfigure}
	\begin{subfigure}{0.255\textwidth}
		\centering
		\includegraphics[width=\textwidth]{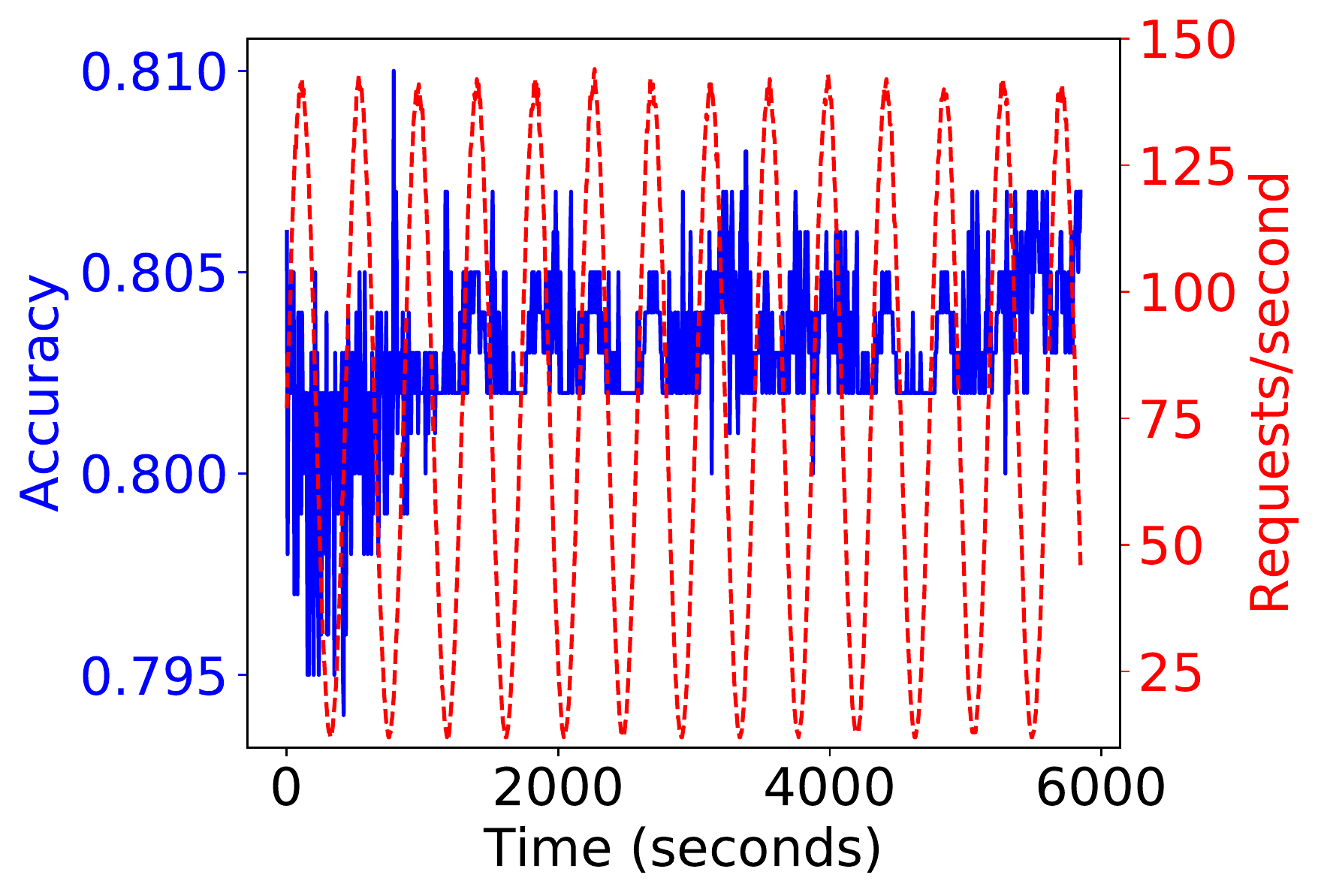}
		\caption{$\beta=1$. \label{fig:acc-sync1a}}
	\end{subfigure}
	\begin{subfigure}{0.235\textwidth}
		\includegraphics[width=\textwidth]{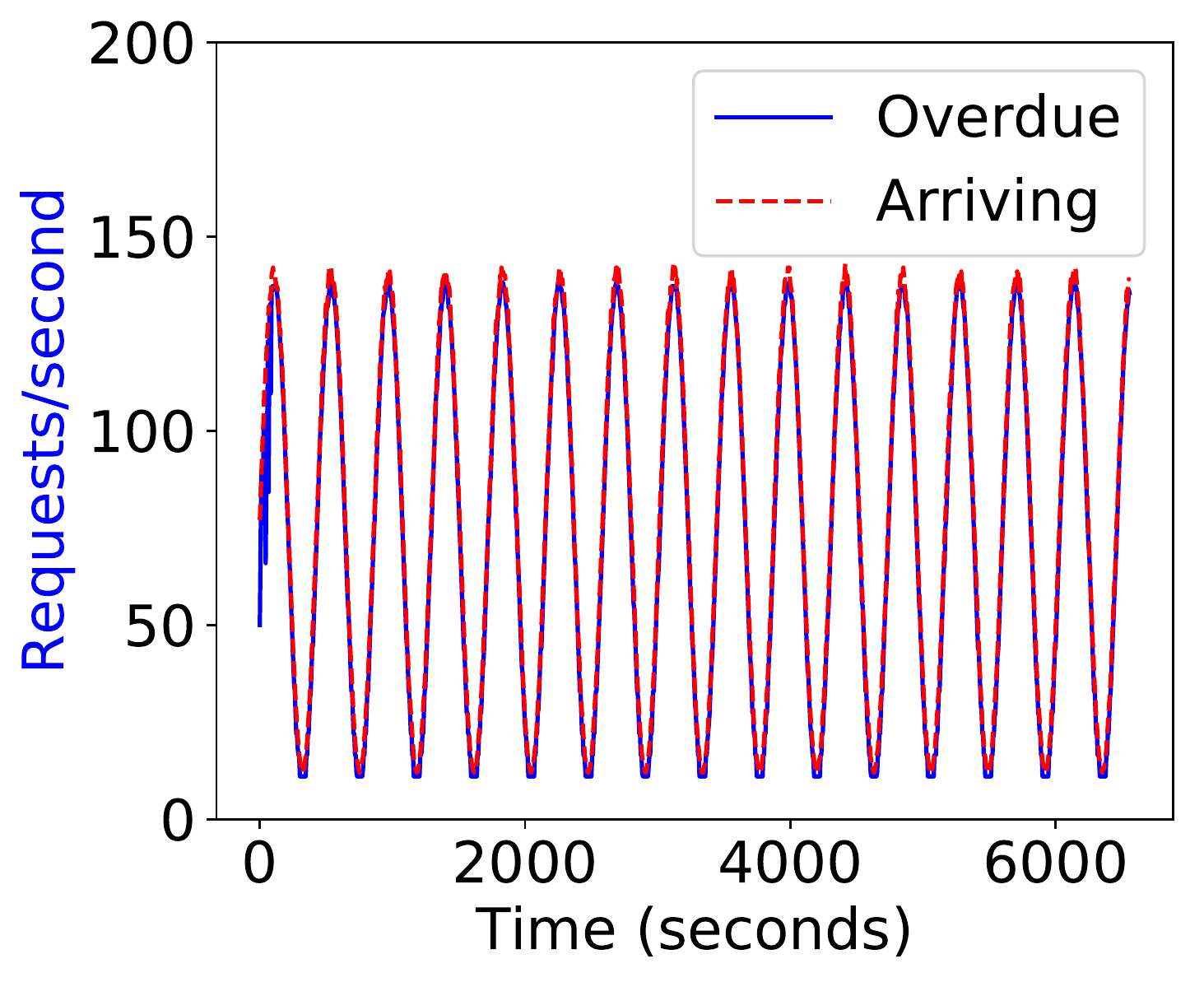}
		\caption{$\beta=0$. \label{fig:overdue-sync0}}
	\end{subfigure}
	\begin{subfigure}{0.235\textwidth}
		\includegraphics[width=\textwidth]{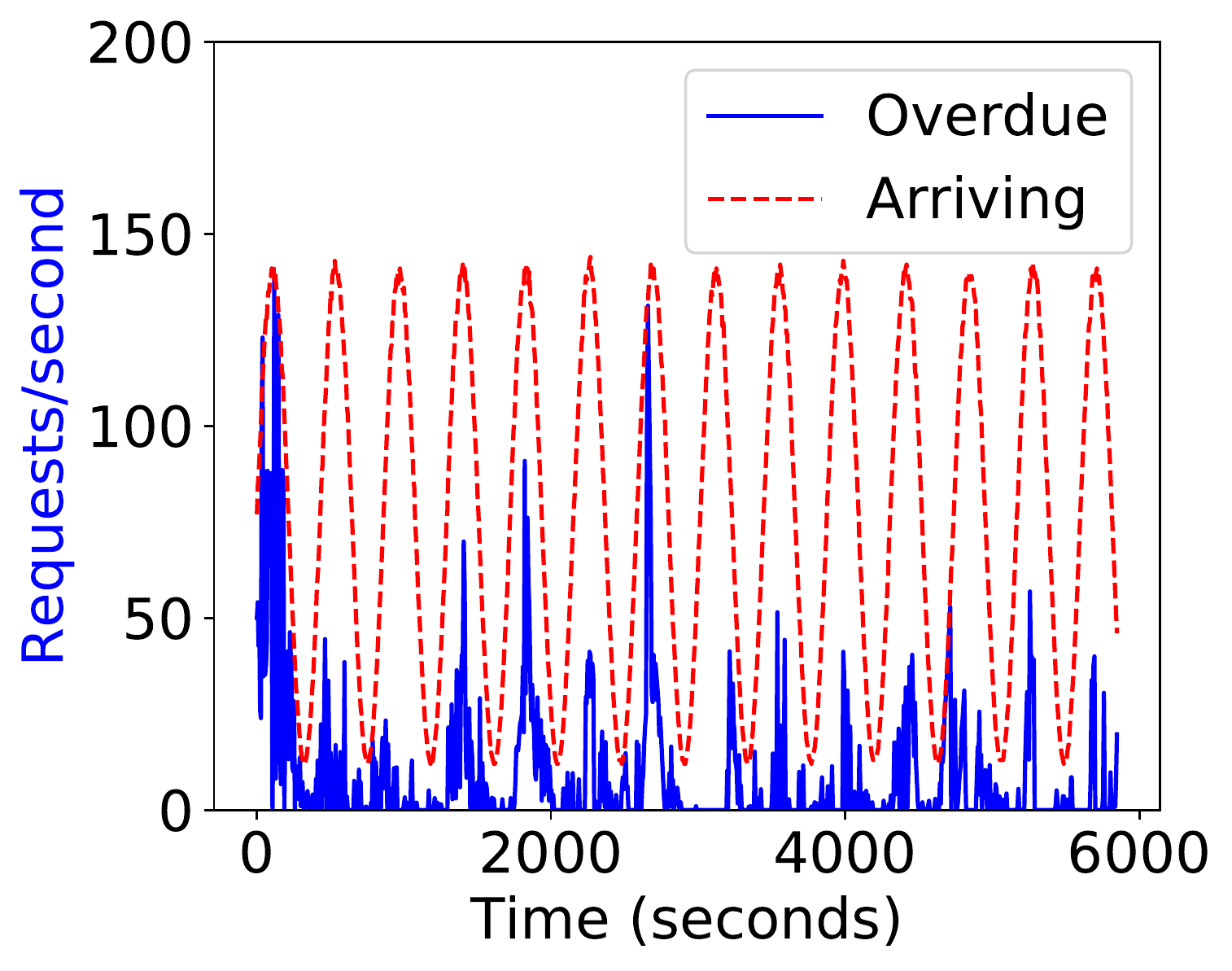}
		\caption{$\beta=1$. \label{fig:overdue-sync1a}}
	\end{subfigure}
	\caption{Comparison of the effect of different $\beta$ for the RL algorithm.\label{fig:beta}}
\end{figure*}

Our environment simulator randomly samples images from the validation dataset as requests. The number of requests is determined as follows. First, to simulate the scenario with very high arriving rate and very low arriving rate, we set the arriving rate based on the maximum throughput $r_u$ and minimum throughput $r_l$. Specifically, the maximum throughput is the sum of all models' throughput, which is achieved when all models run asynchronously to process different batches of requests. In contrast, when all models run synchronously, the slowest model's throughput is just the minimum throughput. Second, we use a sine function and the extreme throughput values to define the request arriving rate, i.e. $r=\gamma sin(t)+b$. The slope $\gamma$ and intercept $b$ are derived by solving Equation~\ref{eq:r1} and \ref{eq:r2}, where T is the cycle period which is configured to be $500\times \tau$ in our experiments. The first equation is to make sure that more requests than $r_u$ (or $r_l$) are generated for 20\% of each cycle period (Figure~\ref{fig:sine}, $T\times 20\%=0.2T$). 
In this way, we are simulating the real-life environment where there are overwhelming requests coming at times. The second equation is to make sure that the highest arriving rate is not too large, otherwise the request queue would be filled up very quickly and new requests have to be dropped. 
Finally, a small random noise is applied over $r$ to prevent the RL algorithm from remembering the $sine$ function.  To conclude, the number of new requests is $\delta \times (\gamma sin(t)+b)\times (1 + \phi), \phi\sim\mathcal N(0, 0.1))$, where $\delta$ is the time span between the last invocation of the simulator and the current time.

\begin{eqnarray}
k\times sin(T/2-0.2\times 2T/2) + b = r_u\ \text{or}\ r_l \label{eq:r1}\\
k\times sin(T/2) + b = 1.1\times r_u\ \text{or}\ r_l \label{eq:r2}
\end{eqnarray}

\subsubsection{Single Inference Model}\label{sec:single-exp}

We use inception\_v3 trained over ImageNet as the single inference model. Besides the greedy algorithm (Algorithm~\ref{alg:greedy}), we also run the RL algorithm from Section~\ref{sec:multiple} to decide the batch size. 
The state is the same as that in Section~\ref{sec:multiple} except that the model related status is removed.

The batch size list is $B=\{16,32,48,64\}$. The maximum throughput is $\max b/c(b)=64/0.23=272$ images/second and the minimum throughput is $\min b/c(b)=16/0.07=228$. We set $\tau=c(64)\times 2=0.56$s. Figure~\ref{fig:single-max} compares the RL algorithm and the greedy algorithm with the arriving rate defined based on the maximum throughput $r_u$. We can see that after a few iterations, RL performs similarly as the greedy algorithm when the request arriving rate $r$ is high. When $r$ is low, RL performs better than the greedy algorithm. This is because there are a few requests left when the queue length does not equal to any batch size in Line 7 of Algorithm~\ref{alg:greedy}. These left requests are likely to overdue because the new requests are coming slowly to form a new batch. Figure~\ref{fig:single-min} compares the RL algorithm and the greedy algorithm with the arriving rate defined based on the minimum throughput $r_l$. We can see that RL performs better than the greedy algorithm when the arriving rate is either high or low. Overall, since the arriving rate is smaller than that in Figure~\ref{fig:single-max}, there are fewer overdue requests.

\subsubsection{Multiple Inference Models}\label{sec:multiple-exp}

In the following experiments, we select inception\_v3, inception\_v4 and inception\_resnet\_v2 to construct the model list $M$. 
The maximum throughput and minimum throughput are 572 requests/second and 128 requests/second respectively. The experimental results are plotted in Figure~\ref{fig:sync}, Figure~\ref{fig:async} and Figure~\ref{fig:beta}. The legend text `Overdue' represents for the number of overdue requests per second, and `Arriving' stands for the request arriving rate.

We compare our RL algorithm with two baseline algorithms respectively. For each baseline, we use the greedy algorithm to find the batch size. They differ in the model selection strategy. The first baseline runs all models synchronously for each batch of requests. Correspondingly, the requests are generated using the rate based on $r_l$. From Figure~\ref{fig:acc-sync} and Figure~\ref{fig:acc-sync1}, we can see that the greedy algorithm has the fixed accuracy, whereas the RL algorithm's accuracy is high (resp. low) when the rate is low (resp. high). In fact, the synchronous algorithm always uses all models to do ensemble modelling, whereas the RL algorithm uses fewer models to do ensemble modeling when the arriving rate is high. Since the request arriving rate is very low, the baseline is able to handle almost all requests. Similar to Figure~\ref{fig:single-min}, some overdue requests in Figure~\ref{fig:overdue-sync} are due to the mismatch of the queue size and the batch size in Algorithm~\ref{alg:greedy}. 

The second baseline runs all models asynchronously, one model per batch of requests. In other words, there is no ensemble modeling.  Correspondingly, the requests are generated using the rate based on $r_u$. We can see that RL has better accuracy (Figure~\ref{fig:acc-async} and \ref{fig:acc-async1}) and fewer overdue requests (Figure~\ref{fig:overdue-async} and \ref{fig:overdue-async1}) than the baseline. Moreover, it is adaptive to the request arriving rate. When the rate is high, it uses fewer models to serve the same batch to improve the throughput and reduce the overdue requests. When the rate is low, it uses more models to serve the same batch to improve the accuracy. 

We also compare the effect of different $\beta$, namely $\beta=0$ and $\beta=1$ in the reward function, i.e. Equation~\ref{eq:multi_acc_reward}. Figure~\ref{fig:beta} shows the results with the requests generated based on $r_l$. We can see that when $\beta$ is smaller, the accuracy (Figure~\ref{fig:acc-sync0}) is higher. This is because the reward function focus more on the accuracy part. Consequently, there are many overdue requests as shown in Figure~\ref{fig:overdue-sync0}. In contrast, when $\beta=1$, the reward function tries to avoid overdue requests by reducing the number of models for ensemble. Therefore, the accuracy and the number of overdue requests are smaller (Figure~\ref{fig:acc-sync1a} and Figure~\ref{fig:overdue-sync1a}).

\section{Case Study on Usability}
\label{sec:demo}

\begin{figure}[!h]
	\centering
	\includegraphics[width=0.3\textwidth]{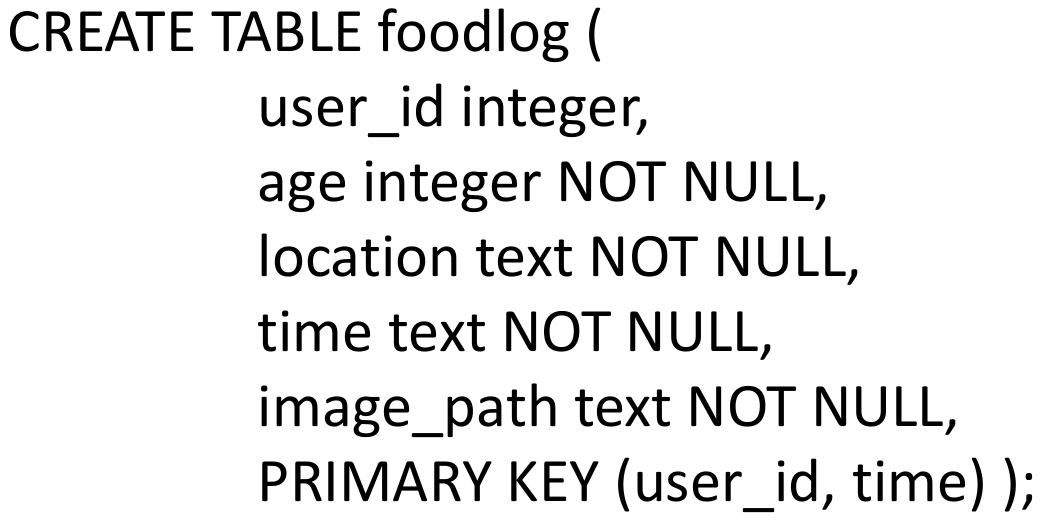}
	\caption{SQL for table creation.\label{fig:sql_table}}
\end{figure}

\begin{figure}[!h]
	\centering
	\includegraphics[width=0.45\textwidth]{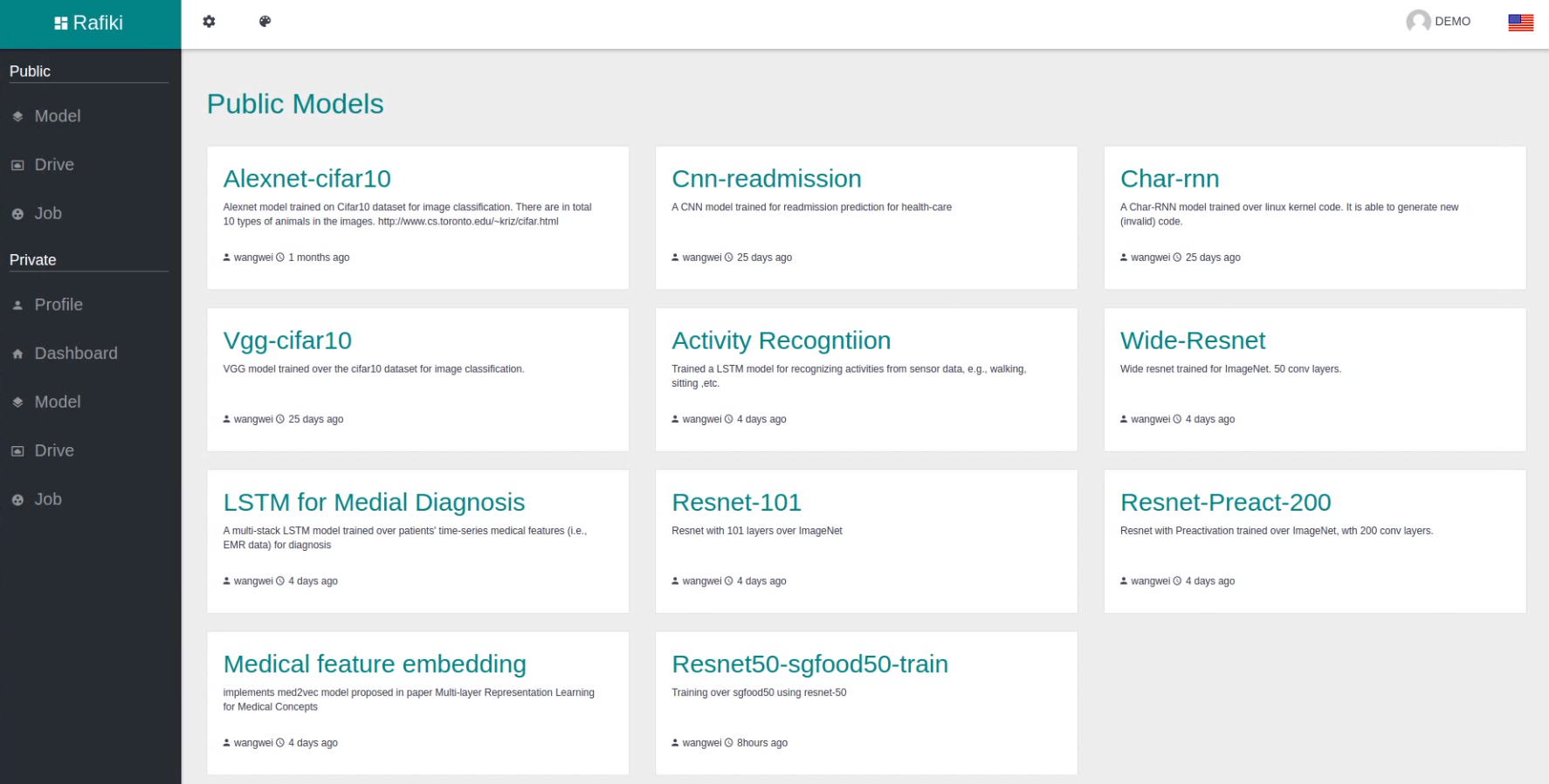}
	\caption{Rafiki web interface.\label{fig:web}}
\end{figure}

In this section, we detail how Rafiki can help a database expert use deep learning services easily in an existing application. 
We present a scenario for analyzing the food preference of users of a food logging application as shown in Figure~\ref{fig:arch}. 
This application keeps a log for the food that the user eats by storing the time, location and photo of each meal. For simplicity, we assume that all data is stored in one table created in Figure~\ref{fig:sql_table}.
Each row in the table contains information about a meal. 
The image\_path is a file path for an image. This image contains the picture of the food, however there is no direct way to query the information (e.g., name) of the food image. Using Rafiki, the database developer collaborates with a deep learning expert who develops and trains an image recognition model for the food images. 
This trained model is shared in Rafiki and provides recognition services to database users as a black box via Web APIs. Figure~\ref{fig:web} displays the web interface of Rafiki.

\textbf{Deep Learning Expert} The deep learning expert prepares the training (train.py) and serving (serve.py) scripts for a deep learning model~\cite{DBLP:journals/corr/HeZRS15} using Apache SINGA~\footnote{http://singa.apache.org/}.

\textbf{Database User} After uploading the training dataset, i.e. a set of food image and label pairs, into the Rafiki, the database user starts the training job. 
Once the training is finished, the user deploys the trained model for a serving job. 
Afterwards, he exploits the deep learning service to get the food name by sending a request to Rafiki via an user-defined function (UDF).
The food\_name() function calls the Web API of the serving application in Rafiki. In particular, the function is executed only on the images of the rows that satisfy the condition (age is greater than 30). 
Consequently, it saves much time. 
Furthermore, when the model is modified or re-trained in Rafiki, there is no change to the SQL query at the database user's side. 

\begin{figure}[!h]
\centering
\includegraphics[width=0.36\textwidth]{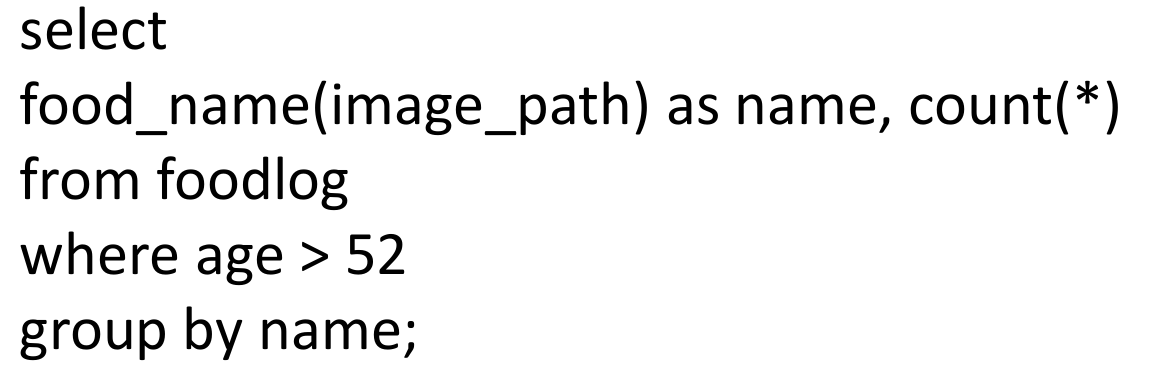}
\end{figure}

This service can also serve requests from mobile Apps as demonstrated in Figure~\ref{fig:arch}. We have created a food logging App which sends requests to Rafiki for food image prediction.

\section{Conclusions}
Complex analytics has become an inherent and expected functionality to be supported by big data systems. However, it is widely recognized that machine learning models are not easy to build and train, and they are sensitive to data distributions and characteristics. 
It is therefore important
to reduce the pain points of implementing and tuning of dataset specific models, as a step towards making AI more usable.
In this paper, we present Rafiki to provide the training and inference service of machine learning models, and facilitate complex analytics on top of cloud data storage systems.  
Rafiki supports effective distributed
hyper-parameter tuning for the training service, and online ensemble  modeling  for  the  inference  service  that is amenable to the trade  off  between latency and accuracy.  
The system is evaluated with various benchmarks to illustrate its efficiency,
effectiveness and scalability.
We also conducted a case study that demonstrates how the system enables a database developer to use deep learning services easily.

\bibliographystyle{abbrv}
\bibliography{dlaas}  

\end{document}